\documentclass[review]{elsarticle}

\usepackage{hyperref}
\usepackage{amsmath}    
\usepackage{amsfonts}
\usepackage{color}          
\usepackage{longtable}    
\usepackage{aas_macros} 
\usepackage{multicol}
\usepackage{multirow}
\usepackage{xtab}
\usepackage{booktabs}
\usepackage{array}
\usepackage{bm}

\usepackage{caption}
\DeclareCaptionLabelFormat{continufloat}{#1 #2\Alph{ContinuedFloat}} 

\journal{Icarus}

\biboptions{authoryear}

\makeatletter
\def\ps@pprintTitle{%
    \let\@oddhead\@empty
    \let\@evenhead\@empty
    \def\@oddfoot{\footnotesize\itshape
         { } \hfill\today}%
    \let\@evenfoot\@oddfoot
    }
\makeatother

\begin{document}

\begin{frontmatter}

\title{Observations and numerical modelling of a convective disturbance in a large-scale cyclone in Jupiter's South Temperate Belt}



\author[address1]{P. I\~nurrigarro\corref{correspondingauthor}}
\ead{peio.inurrigarro@ehu.eus}
\author[address1]{R. Hueso}
\author[address1]{J. Legarreta}
\author[address1]{A. S\'anchez-Lavega}
\author[address2]{G. Eichst\"adt}
\author[address3]{J. H. Rogers}
\author[address4]{G. S. Orton}
\author[address5]{C. J. Hansen}
\author[address1]{S. P\'erez-Hoyos}
\author[address1]{J. F. Rojas}
\author[address6]{J. M. G\'omez-Forrellad}
\cortext[correspondingauthor]{Corresponding author}
\nonumnote{Paper published as: P. I\~nurrigarro et al., 2020. Observations and numerical modelling of a convective disturbance in a large-scale cyclone in Jupiter's South Temperate Belt. Icarus 336. DOI: \url{https://doi.org/10.1016/j.icarus.2019.113475}.}

\address[address1]{Fisica Aplicada I, Escuela de Ingenieria de Bilbao, UPV/EHU, Plaza Ingeniero Torres Quevedo, 1 48013 Bilbao, Spain}
\address[address2]{Independent scholar, Stuttgart, Germany}
\address[address3]{British Astronomical Association, Burlington House, Piccadilly, London W1J 0DU, UK}
\address[address4]{Jet Propulsion Laboratory, California Institute of Technology, 4800 Oak Grove Drive, Pasadena, CA 91109, USA}
\address[address5]{Planetary Science Institute, 1700 East Fort Lowell, Suite 106, Tucson, AZ 85719-2395, USA}
\address[address6]{Fundaci\'o Observatory Esteve Duran, Barcelona, Spain}

\begin{abstract}
Moist convective storms in Jupiter develop frequently and can trigger atmospheric activity of different scales, from localized storms to planetary-scale disturbances including convective activity confined inside a larger meteorological system. In February 2018 a series of convective storms erupted in Jupiter's South Temperate Belt (STB) (planetocentric latitudes from $-23^{\circ}$ to $-29.5^{\circ}$). This occurred inside an elongated cyclonic region known popularly as the STB Ghost, close to the large anticyclone Oval BA, resulting in the clouds from the storms being confined to the cyclone. The initial storms lasted only a few days but they generated abundant enduring turbulence. They also produced dark features, possibly partially devoid of clouds, that circulated around the cyclone over the first week. The subsequent activity developed over months and resulted in two main structures, one of them closely interacting with Oval BA and the other one being expelled to the west. Here we present a study of this meteorological activity based on daily observations provided by the amateur community, complemented by observations obtained from PlanetCam UPV/EHU at Calar Alto Observatory, the Hubble Space Telescope and by JunoCam on the Juno spacecraft. We also perform numerical simulations with the EPIC General Circulation Model to reproduce the phenomenology observed. The successful simulations require a complex interplay between the Ghost, the convective eruptions and Oval, and they demonstrate that water moist convection was the source of the initial storms. A simple scale comparison with other moist convective storms that can be observed in the planet in visible and methane absorption band images strongly suggests that most of these storms are powered by water condensation instead of ammonia.
\end{abstract}


\end{frontmatter}

\section{Introduction}

Moist convective storms in Jupiter are frequent and vigorous and they are suspected to play a key role in the planet's meteorology \citep{Ingersoll_2000, Vasavada_2005}. Some of these storms can trigger large-scale changes in Jupiter's atmosphere, changing the visual aspect of a belt or zone and developing planetary-scale perturbations. The most relevant cases occur in the South Equatorial Belt (SEB) \citep{Sanchez-Lavega_Icarus_1996, Perez-Hoyos_2012_SEB_2010_fade, Fletcher_2011_SEB_fade, Fletcher_2017_SEB_storms} and the North Temperate Belt (NTB) \citep{Sanchez-Lavega_Nature_NTBD_2008, Sanchez-Lavega_NTBD_2017, Barrado_2009_NTBD_2007}. However, most convective storms are limited in size and temporal duration, like the common storms found in the northwest wake of the Great Red Spot (GRS), and seem not to have important consequences on their environment. This might be due in part to the latitudinal shear of the wind that disperses the storm material in time scales of a few days \citep{Hueso_JGR_2002}. On Jupiter, convective storms develop preferentially in the cyclonic side of the zonal jets or in regions of enhanced cyclonicity and are rarely observed forming in other environments \citep{Little_Lightning_1999, Ingersoll_2004, Vasavada_2005}. \citet{Dowling_1989_Cyclones_convection_Jupiter} proposed that in the jovian atmosphere cyclones should cause a depression in surfaces of equal potential temperature in the upper layers and rises in the deep layers that could serve to start moist convection in the water cloud layer.

Since the arrival of the Juno mission to Jupiter in July 2016, the giant planet has been observed more frequently than in any previous period of time. Recent advancements in fast-acquisition cameras and in image-processing software allow observers to obtain images with small telescopes that are significantly better than a decade ago (see \citealt{Mousis_2014_PRO-AM} for a review of modern amateur methods). The combination of modern Jupiter observations by many amateur astronomers results in a daily survey of its atmospheric activity during most of the year \citep{Hueso_2018_PVOL}. Among the many topics on Jovian meteorology that can be characterized with such an observational coverage, one is the fast development and long-term evolution of convective eruptions in the planet.

\begin{figure}[htbp]
\centering
\includegraphics[angle=0, width=1.0\textwidth]{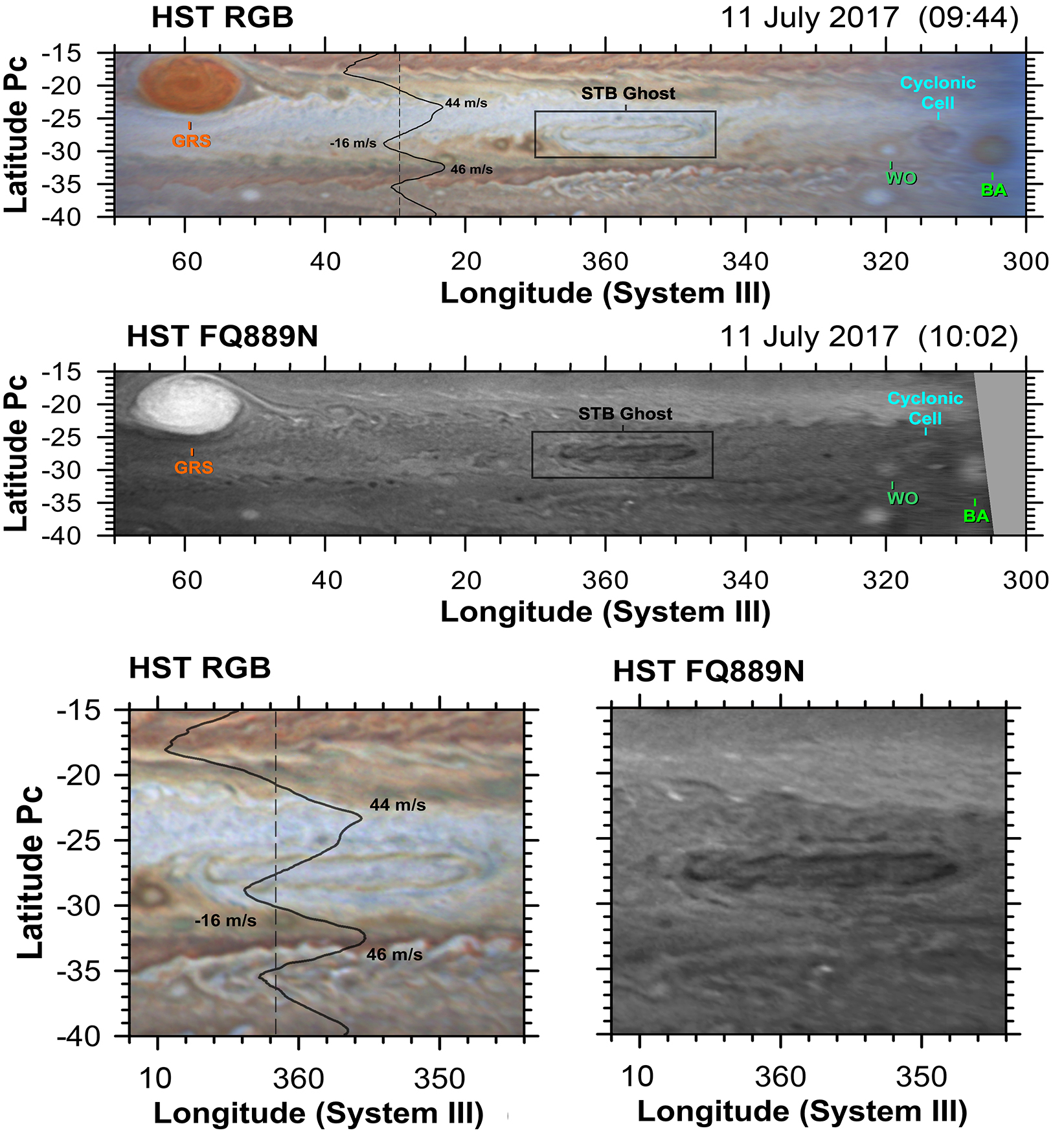}
\caption{Cylindrical maps of Jupiter's South Temperate Belt from images acquired with the Hubble Space Telescope (HST) on 11 July 2017. The maps show the STB Ghost before the development of the convective storms. The upper image is a colour composite map of the STB jet and its environment (F631N, F502N and F395N filters were used as the RGB channels), the middle image was taken with the FQ889N filter, at a strong methane absorption band centred at $889\,$nm showing details of the upper atmosphere. The bottom row shows close-ups of the STB Ghost in the visible (left) and strong methane band (right). The images have been processed using high-pass filters and combined taking into account planetary rotation to enhance the visibility of small-scale features. The positions of the STB Ghost, the large anticyclone Oval BA, a cyclonic cell northwest of Oval BA and a white oval (WO) are highlighted in the upper two images. The background zonal-wind profile from 2016 HST observations \citep{Hueso_Jupiter_before_Juno_2017} is also shown on the upper and bottom left panels, the dashed line represents the null zonal-wind velocity.}
\label{figure HST Ghost before storms}
\end{figure}

On 4 February 2018 a series of convective storms erupted in the western side of a low-contrast elongated cyclone in the South Temperate Belt (STB) of the planet, which extends between planetocentric latitudes $-23^{\circ}$ and $-29.5^{\circ}$. The cyclone was located at  planetocentric latitude $-27.5^{\circ}$ and formed in 2013. The early history of this cyclone, called the STB Ghost largely due to its bland, colourless appearance in ground-based observations, is given by \citet{Rogers_2015, Rogers_2019}. The Hubble Space Telescope (HST) observed the STB Ghost several times before the convective eruptions. Figure \ref{figure HST Ghost before storms} shows two maps of the Ghost from HST images acquired in July 2017 before the convective eruptions of February 2018. The Ghost is a noticeable dark feature in images acquired in the methane band showing less upper-cloud material. Elongated cyclonic features are not unusual in Jupiter's STB, and similar structures were observed at the same latitude at the time of the Voyager 1, Voyager 2, and Cassini fly-bys and in ground-based observations during 2004-2009 and from 2016 onwards \citep{Rogers_2015, Rogers_2019}. All these cyclones are pale compared with the similarly elongated red and dark cyclones known as barges, that have been observed in the North Equatorial Belt (NEB) at around planetocentric latitude $14^{\circ}$ \citep{Hatzes_Voyager_Barge_1981, Legarreta_Jupiter_cyclones_anticyclones_2005} and during the faded state of the SEB between planetocentric latitudes $-14^{\circ}$ and $-16^{\circ}$. Barges in the SEB have also been home to the start of large convective events \citep{Hueso_JGR_2002} that can grow and impact the life cycle of the SEB \citep{Fletcher_2011_SEB_fade}. 

\begin{figure}[h]
\centering
\includegraphics[angle=0, width=1.0\textwidth]{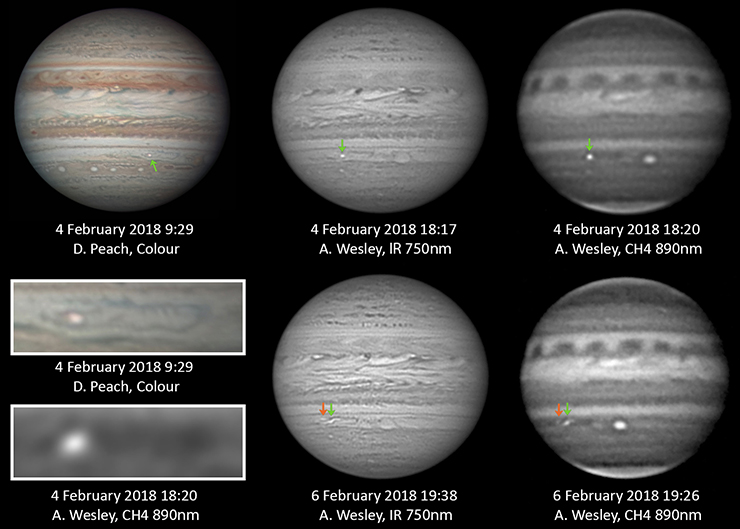}
\caption{Onset of the convective activity that developed the South Temperate Belt Disturbance. All the observations were taken by amateur astronomers identified in the figure legends. The locations of the outbreaks have been marked with a green arrow for the first storm and with an orange-red arrow for the second one.}
\label{figure onset storms}
\end{figure}

The convective eruption in the STB Ghost and the subsequent evolution were recorded by dozens of amateur observers using small-size telescopes. Figure \ref{figure onset storms} shows observations of the first storm on 4 February 2018 in the visible and in the deep methane absorption band at $890\,$nm. The bright visual aspect of the storm at $890\,$nm probes the high-cloud tops associated with the developing bright cloud. Previous observations of the Ghost in the visible obtained the day before do not show this bright cloud, demonstrating a fast development. Later observations show the development of an additional convective core and a possible third one in the course of a few days. The location of these storms inside the cyclonic Ghost resulted in a confinement of the convective clouds that produced strong turbulence in the Ghost characterized by the evolution and circulation of bright and dark patches. In just a few days the Ghost was fully perturbed but its later evolution extended for months, forming a long-lasting South Temperate Belt Disturbance (STBD) that was confined in longitudes inside the STB Ghost. Due to the exceptional characteristics of Jupiter observations during 2018 supporting the Juno mission, this kind of phenomenon has never been observed before in such detail.

Here, we characterize these storms, and how they transformed the STB Ghost into a large-scale STB Disturbance. We use a combination of observations from the amateur community, the PlanetCam UPV/EHU instrument at Calar Alto Observatory, HST and JunoCam. Additionally, we use the Explicit Planetary Isentropic Model (EPIC) \citep{Dowling_1998} to investigate the properties of the atmosphere and the intensity of convection required to reproduce the observations. For that purpose the model has been modified to include the effects of convection through the addition of a heat impulse following the approach of \citet{Garcia-Melendo_2005}. 

The structure of this paper is as follows. Section 2 presents the observations. Section 3 contains our analysis of the STB Ghost and the evolution of the disturbance. Section 4 details our numerical experiments with the EPIC model to simulate the observed phenomenology. We discuss the results of the simulations in Section 5 and we present our conclusions on Section 6. Unless otherwise expressed, all longitudes will be given in System III and all latitudes will be planetocentric.

\section{Observations} \label{section_observations}

\begin{figure}[htbp]
\centering
\includegraphics[angle=0, width=0.55\textwidth]{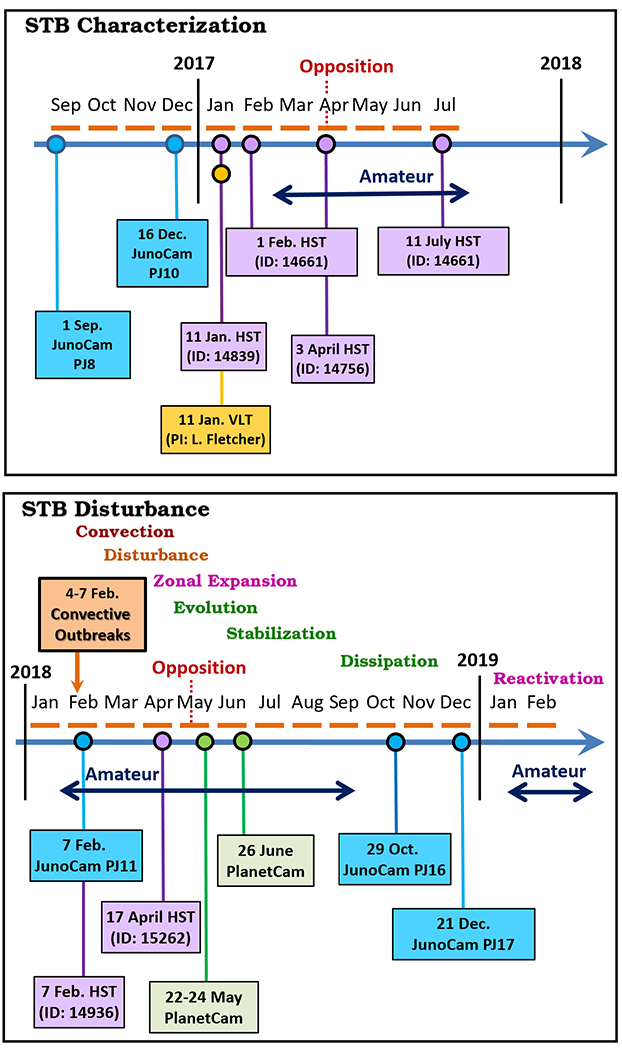}
\caption{Timeline of the observations. The upper panel details the observations employed for the characterization of the STB Ghost and Oval BA. The bottom panel shows the observations used to study the outbreaks and later evolution of the phenomenon. Circles in the timeline indicate specific high-resolution observations from JunoCam (blue), HST (violet), PlanetCam (green) and VLT (dark yellow). Blue arrows indicate periods of time covered by amateur observations. Boxes reference the high-resolution observations used. Different phases of the evolution of the STB Disturbance appear highlighted.}
\label{figure Timeline}
\end{figure}

We here describe the different data sets used grouped by their source. Figure \ref{figure Timeline} shows a timeline of the observations compared with the different phases of the phenomenon. These phases will be discussed in Section \ref{section_analysis}. 

\subsection{Amateur observations}

We used selected images obtained by amateur astronomers from 25 January 2018 to 13 September 2018 and retrieved from the PVOL database (Planetary Virtual Observatory and Laboratory, \url{http://pvol2.ehu.eus}, \citealt{Hueso_2010_PVOL, Hueso_2018_PVOL}) and ALPO-Japan (Association of Lunar and Planetary Observers in Japan, \url{http://alpo-j.asahikawa-med.ac.jp/indexE.htm}). About 190 high-resolution observations covering this period were used. A small selection of additional images obtained over 2017 and 2019 were used to determine the drift rates of the STB Ghost and other nearby features such as the large Oval BA before the onset of the perturbation and to study the outcome of the disturbance. Table \ref{table_amateur} in Appendix A details the observations from the amateur community used here including names of the individual observers. Figure \ref{figure onset storms} shows representative examples of these observations.

\subsection{Calar Alto PlanetCam observations}

The PlanetCam UPV/EHU instrument \citep{Mendikoa_2016_PlanetCam} is a fast-acquisition dual camera developed to obtain high-resolution images of Solar System planets using the lucky-imaging technique \citep{Law2006}. The instrument observes at two wavelength ranges simultaneously with two independent cameras that  constitute two channels: the visible from $0.38\, \mu$m to $1\, \mu$m, and the short-wave infrared (SWIR), from $1\, \mu$m to $1.7\, \mu$m. The instrument has a series of narrow-band and wide-band filters going from colour to methane absorption bands and their adjacent continuum. We ran two observing campaigns at Calar Alto Observatory using the $2.2\,$m telescope in May and June 2018 in which we obtained good observations of the STBD.

\begin{table}[h]
 \centering
 \begin{tabular}{|l|l|l|l|l|}
  \hline
\textbf{Date}   & \textbf{Time} & \textbf{Filter} &\textbf{Central Wavelength} & \textbf{Filter Width} \\
(yyyy/mm/dd)  & (UT)              &                     &   (nm)                              & (nm)          \\
\hline
  2018/05/22    & 00:41       & Bessel I     & 880  & 289     \\ 
  2018/05/22    & 20:36       & Bessel I     & 880  & 289     \\
  2018/05/22    & 20:36       & H            & 1630 & 300     \\
  2018/05/22    & 20:50       & M3            & 890  & 5       \\ 
  2018/05/24    & 21:36       & RG1000        & ---  & $1\, \mu$m to $1.7\, \mu$m \\
  2018/05/24    & 22:40       & Bessel I     & 880  & 289     \\ 
 \hline
  2018/06/23    & 23:23       & Bessel I     & 880  & 289     \\ 
 \hline
 \end{tabular}
\caption{PlanetCam UPV/EHU observations used in this study. }
\label{tabla planetcam}
\end{table}

Table \ref{tabla planetcam} summarizes the characteristics of the observations used for this work. Observations in many other filters available in the PlanetCam instrument were also acquired and will be used in the future for our ongoing analysis of changes in the cloud properties of the planet (see \citealt{Mendikoa_2017_PlanetCam}). 

Images in the Bessel I filter achieved the highest resolution in this data set. This was caused by the smaller impact of atmospheric seeing on the image quality at long wavelengths and the higher sensitivity of the visible detector in the PlanetCam instrument when compared with the detector in the SWIR channel. Photons in the wavelength range of the Bessel I filter penetrate deep in the jovian atmosphere in the absence of clouds and, thus, observations in the I filter are mainly sensitive to the opacity of the main cloud observed in visible wavelengths. Observations in the $890$-nm filter have their peak sensitivity in Jupiter for a cloud-free atmosphere around $200\,$mbar and are assumed to sense the distribution of upper hazes (see for instance, \citealt{West2004}). Bright compact features in this filter and in the visible are generally considered a good signature of vertically extended convective clouds \citep{Sanchez-Lavega_Nature_NTBD_2008}. Among the SWIR filters, the highest resolution was obtained with the RG1000 (1.0-1.7 $\mu$m) and H-band filters. These filters are very broad and have vertically extended contribution functions that make them sensitive to both the upper hazes around the tropopause and the deep cloud, which is more likely the most important contribution. Vertically extended convective clouds would appear bright in all these filters.

\begin{figure}[htbp]
\centering
\includegraphics[angle=0, width=1.0\textwidth]{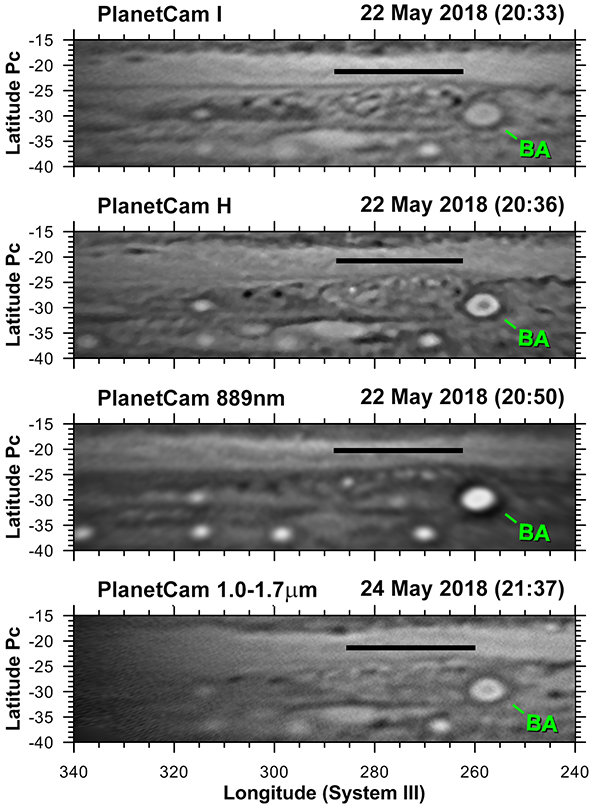}
\caption{Maps of the region of interest from PlanetCam observations in May 2018. The images show the structure of the perturbed region about 3 months after the beginning of the disturbance. Filters are annotated in each map. A line marks the position of the disturbed STB Ghost. Oval BA is also highlighted. Features as small as 500 km are resolved in these images.}
\label{figure PlanetCam}
\end{figure}

These images were produced by processing videos of 1-minute length with the instrument pipeline PLAYLIST (PLAnetarY Lucky Images STacker). Several consecutive videos (up to 10 videos in the observations aimed for the highest resolution) were obtained in the Bessel I, the $890\,$nm wavelength, the H band and the RG1000 filters and were navigated and combined into single images through a derotation process run with the WinJUPOS software (\url{http://jupos.privat.t-online.de/index.htm}). This lengthy process is essentially the same one used by most amateurs to obtain their best observations. Examples of maps of the region of interest of the final images after processing are shown in Figure \ref{figure PlanetCam}.

\subsection{HST images}

\begin{table}[h]
 \centering
 \begin{tabular}{|l|l|l|}
  \hline
  \textbf{Date}   & \textbf{Filters} & \begin{tabular}[c]{@{}l@{}}\textbf{HST program} \end{tabular} \\ 
 (yyyy/mm/dd)   &                      &  \begin{tabular}[c]{@{}l@{}}\textbf{ID}                \end{tabular} \\  \hline
  2017/01/11      & \begin{tabular}[c]{@{}l@{}}F631N, F395N, F502N, F343N, F275W,  \\ F225W, FQ889N, FQ750N, FQ727N \end{tabular}   & 14839 \\ \hline
  2017/02/01-02  & \begin{tabular}[c]{@{}l@{}}F631N, F395N, F502N, F343N, F275W,  \\ F225W, FQ889N, FQ750N, FQ727N \end{tabular}   & 14661 \\ \hline
  2017/04/03      & \begin{tabular}[c]{@{}l@{}}F631N, F502N, F395N, F467M,             \\ FQ889N, F658N, F275W, F547M   \end{tabular}     & 14756 \\ \hline
  2017/07/11      & \begin{tabular}[c]{@{}l@{}}F631N, F275W, F225W, F395N, F502N,  \\ F343N, FQ727N, FQ750N, FQ889N \end{tabular}   & 14661 \\ \hline
  2018/02/07      & \begin{tabular}[c]{@{}l@{}}FQ889N, FQ727N, F631N, F502N,         \\ F395N, F275W, F225W, F343N    \end{tabular}      & 14936 \\ \hline
  2018/04/17      & \begin{tabular}[c]{@{}l@{}}F631N, F502N, F395N, F467M,             \\ FQ889N, F658N, F275W, F547M   \end{tabular}     & 15262 \\ \hline
 \end{tabular}
\caption{List of HST observations used.}
\label{tabla HST}
\end{table}

HST observations of Jupiter obtained over 2017 and 2018 have been retrieved from the Mikulski Archive for Space Telescopes at \url{https://archive.stsci.edu/}. These observations correspond to the Outer Planets Atmospheres Legacy program (OPAL) \citep{Simon_2015_OPAL} and HST programs related to the Juno mission, including the Wide Field Coverage for Juno (WFCJ) program (\url{https://archive.stsci.edu/prepds/wfcj/}). Observations in 2017 were used to characterize the STB Ghost before the onset of the storms. Observations in 2018 covered different stages of the activity of the Ghost. Table \ref{tabla HST} provides a list of the HST programs, dates of observations and filters available in each HST set.

\subsection{JunoCam}

\begin{figure}[htbp]
	\centering
		\includegraphics[angle=0, width=1.0\textwidth]{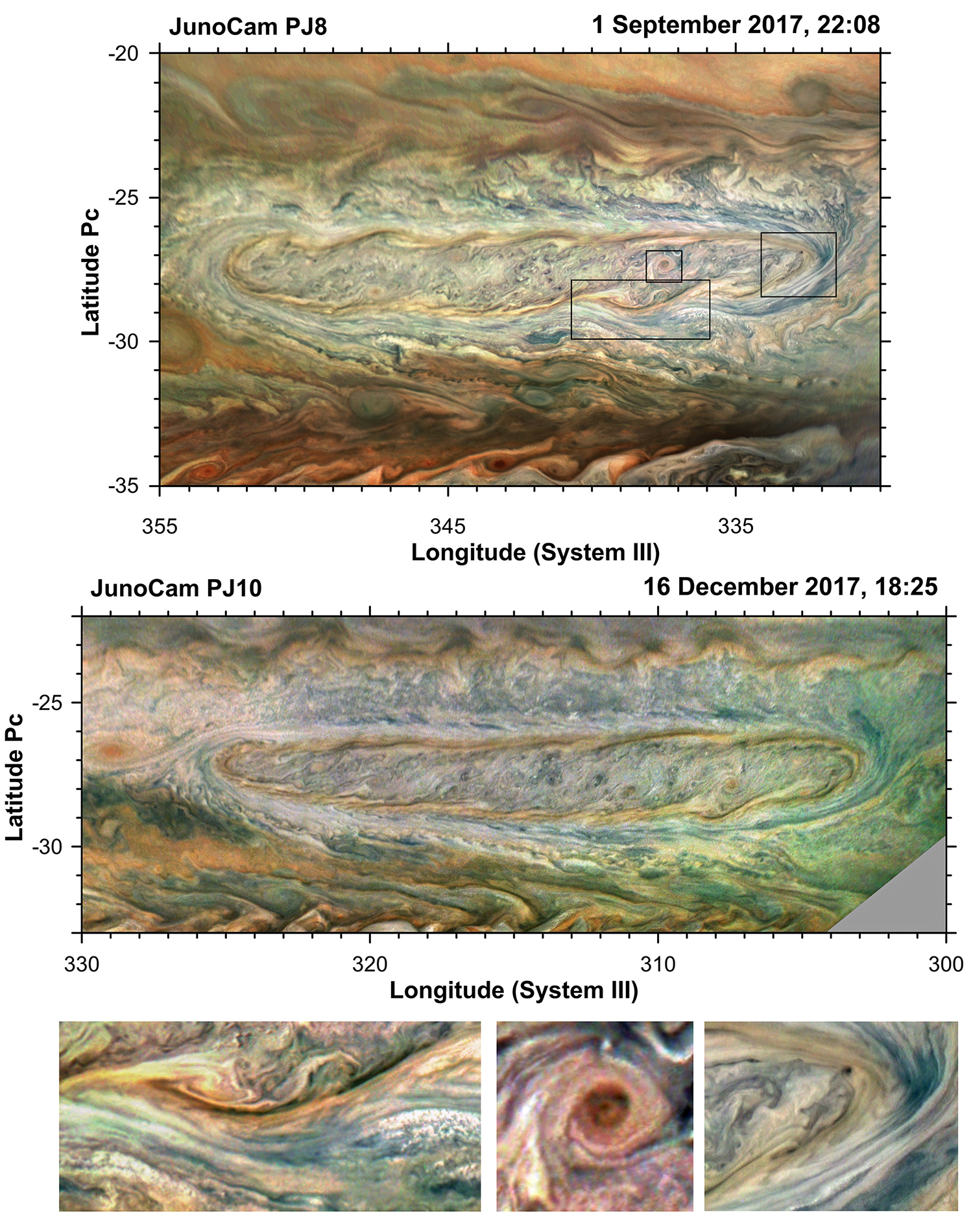}
	\caption{Cylindrical maps of JunoCam images showing the very fine-scale structure of the STB Ghost obtained during the 8\textsuperscript{th} and 10\textsuperscript{th} perijoves on 1 September 2017 and 16 December 2017, respectively. Close-ups of some features of interest in the first image are shown on the bottom row.}
	\label{figure JunoCam}
\end{figure}

JunoCam is a wide-field pushframe imaging instrument on board Juno whose primary goal is public outreach. JunoCam is obtaining some of the most detailed images ever seen of Jupiter's clouds and its analysis is being done in collaboration with citizen scientists that process the images, and amateur astronomers that provide global context to JunoCam images \citep{Hansen_2017_JunoCam}. The performance of JunoCam and scientific use of JunoCam is further detailed by \citet{Orton_2017_JunoCam} and \citet{Sanchez-Lavega_2018_GRS_JunoCam}. We have used two very high-resolution images taken during the 10\textsuperscript{th} perijove on 16 December 2017 to study the dynamics of the STB Ghost and estimate the wind speeds. Further observations of the STB Ghost at varying resolutions were acquired in different perijoves and are described in Table \ref{tabla JunoCam}. Figure \ref{figure JunoCam} shows the best observations of the STB Ghost, which were acquired on the 8\textsuperscript{th} and 10\textsuperscript{th} perijoves on 1 September 2017 and 16 December 2017 respectively. 

JunoCam images were map projected using information from Juno's trajectory from SPICE kernels, Juno's spin, JunoCam's mounting and Jupiter's shape and rotation. The images were also processed to remove camera distortions and an analysis of repetitive camera blemish and hot pixels was used to remove some obvious image defects. The global illumination was adjusted using an empirical bidirectional reflectance model of Jupiter's cloud reflectivity and images were contrast enhanced and further processed to enhance the contrast and visibility of small features visible in Figure \ref{figure JunoCam} \citep{Eichstadt_EPSC_2017}.  We highlight that the  observations acquired on perijove 11 on 7 February 2018 were taken three days after the onset of the convective activity. Also observations on perijoves 16 (29 October 2018) and 17 (21 December 2018) were acquired at dates close to solar conjunction (26 November 2018) with Jupiter and the Sun at distances of $\sim$20$^{\circ}$, making ground-based observations difficult. Thus, JunoCam images were also very valuable to understand the long-term evolution of the disturbance.

\begin{table}[h]
 \centering
 \begin{tabular}{|l|l|l|}
  \hline
  \textbf{Perijove} & \textbf{Date} (yyyy/mm/dd)  & \textbf{Spatial resolution}\\ 
\hline
      5             &   2017/03/27   & $500\,$km \\  
      6             &   2017/05/19   & $500\,$km \\  
      7             &   2017/07/11   & $500\,$km \\  
      8             &   2017/09/01   & $18\,$km \\  
      9             &   2017/10/24   & $1,000\,$km \\  
     10             &   2017/12/16   & $24\,$km \\  
     11*            &   2018/02/07   & $160\,$km  \\ 
     16             &   2018/10/29   & $100\,$km \\  
     17             &   2018/12/21   & $100\,$km \\  \hline
 \end{tabular}
\caption{List of JunoCam perijoves where the STB Ghost or its remains after the STB Disturbance were observed. \textbf{Notes:} The spatial resolution of JunoCam images  varies over individual images. Spatial resolutions in the table are estimates based in the sizes of the features in the best resolved areas inside the Ghost. (*) Images acquired in perijove 11 were acquired three days after the start of the convection.}
\label{tabla JunoCam}
\end{table}

\subsection{Other observations}

The STB Ghost is also visible in some images of Jupiter in the thermal IR obtained with the VISIR instrument on the VLT and already published (Figure 9 in \citealt{Fletcher_2018_NEB_wave}). According to published images at 5 $\mu$m, the STB Ghost appears as a slightly brighter feature compared to the very dark STB with a slightly lower aerosol opacity than the STB but much larger than in the bright and aerosol-free 5-$\mu$m zones. This is consistent with its lower brightness in 890 nm implying slightly less content of high clouds and hazes compared to the regions of the STB surrounding the Ghost.

\section{Analysis} \label{section_analysis}

We have used the WinJUPOS software to navigate the amateur, PlanetCam and HST images and generate cylindrical maps. The maps were corrected from limb darkening effects using a simple Lambert correction that allows the comparison of the morphology of cloud structures observed at different positions from the central meridian of the planet.

The analysis has been divided in four parts: 
\begin{itemize}
\item Ghost characterization: We used cylindrical maps of the STB Ghost prior to the convective eruption from HST and JunoCam images. These images were obtained with time differences of hours and tens of minutes, respectively, allowing us to perform cloud tracking over small features and obtain wind measurements. For this purpose we used the semi-automatic image correlation software PICV (Particle Image Correlation Velocimetry) \citep{Hueso_2009_BA}.  Additionally, we used JunoCam, HST and amateur observations in the months prior to the convective eruptions to characterize the drift rate and size of the STB Ghost before the convective outbreaks.

\item Convective storms: We studied the onset of the convective activity with amateur observations in the visible and in the methane absorption band. We present details of the convective structures based on HST and JunoCam images acquired 3 days after the onset of the first storm.

\item STB Disturbance evolution and dissipation: We used amateur, PlanetCam, HST and JunoCam observations of the STB covering an extended period of time. Amateur images show atmospheric features produced by the activity in the STB Ghost which have been tracked in time over many different images. PlanetCam and HST images give higher resolution images, allowing to explore images in the methane band and search for signatures of possible convection. The final outcome of this evolution has been investigated with JunoCam images obtained in October and December 2018.

\item STB and Oval BA interaction. Since mid-April 2018, the east side of the Ghost and Oval BA came together, resulting in an interaction between both features, which remained close together until the last observations. This interaction was studied using amateur and JunoCam images.
\end{itemize}

\subsection{The STB Ghost before the STBD} \label{section Ghost before}

High-resolution JunoCam observations of the Ghost were acquired on 1 September 2017 and 16 December 2017 (perijoves 8 and 10, respectively). Figure \ref{figure JunoCam} shows these observations with the detailed structure of the Ghost. A variety of dynamical regions and cloud morphologies are visible. Inside the Ghost there are red and compact cyclonic vortices with diameters of $(660-870 \pm 60)\,$km, bright clouds with sizes around $(45-130 \pm 20)\,$km, an undulating collar with a low brown albedo that suggests wave instability with wavelengths in the range of $(3,200 - 4,500 \pm 110)\,$km. Uncertainties here are calculated from the standard deviation of several measurements made for similar features. For small features, such as the bright clouds, the spatial resolution of the images was considered as the uncertainty. This collar is surrounded by a peripheral collar with higher albedo, where there is evidence of a high-speed flow with strong zonal shears of the meridional flow at the east and west edges of the Ghost. Outside the Ghost there are systems of bright small ``puffy'' clouds with sizes as small as $(35\pm 20)\,$km or as large as $(145\pm 20)\,$km.

We obtained an overall estimate of the circulation of the STB Ghost from a pair of JunoCam images obtained on 16 December 2017.  An accurate measurement of the wind field was not possible because the two images were separated by only $12\,$min and $24\,$s, and this small time separation maximizes errors from cloud tracking associated with small navigation errors in any of the two images. We examined the apparent motions of particular details in the west side of the Ghost by blinking between the two JunoCam images. This resulted in an overall estimate of the maximum wind speeds of ($80 \pm 20$)$\,$ms$^{-1}$ in the outer collar, or $50-60\,$ms$^{-1}$ faster than the environment zonal winds. These velocities are larger than those found in the NEB cyclonic barges, where typical velocities measured on Voyager-2 images were $-41$ and $53\,$ ms$^{-1}$ in the north and south limit of the barges respectively, or $21\,$ms$^{-1}$ faster than the environment winds \citep{Hatzes_Voyager_Barge_1981}.

\begin{figure}[h]
\centering
\includegraphics[angle=0, width=0.995\textwidth]{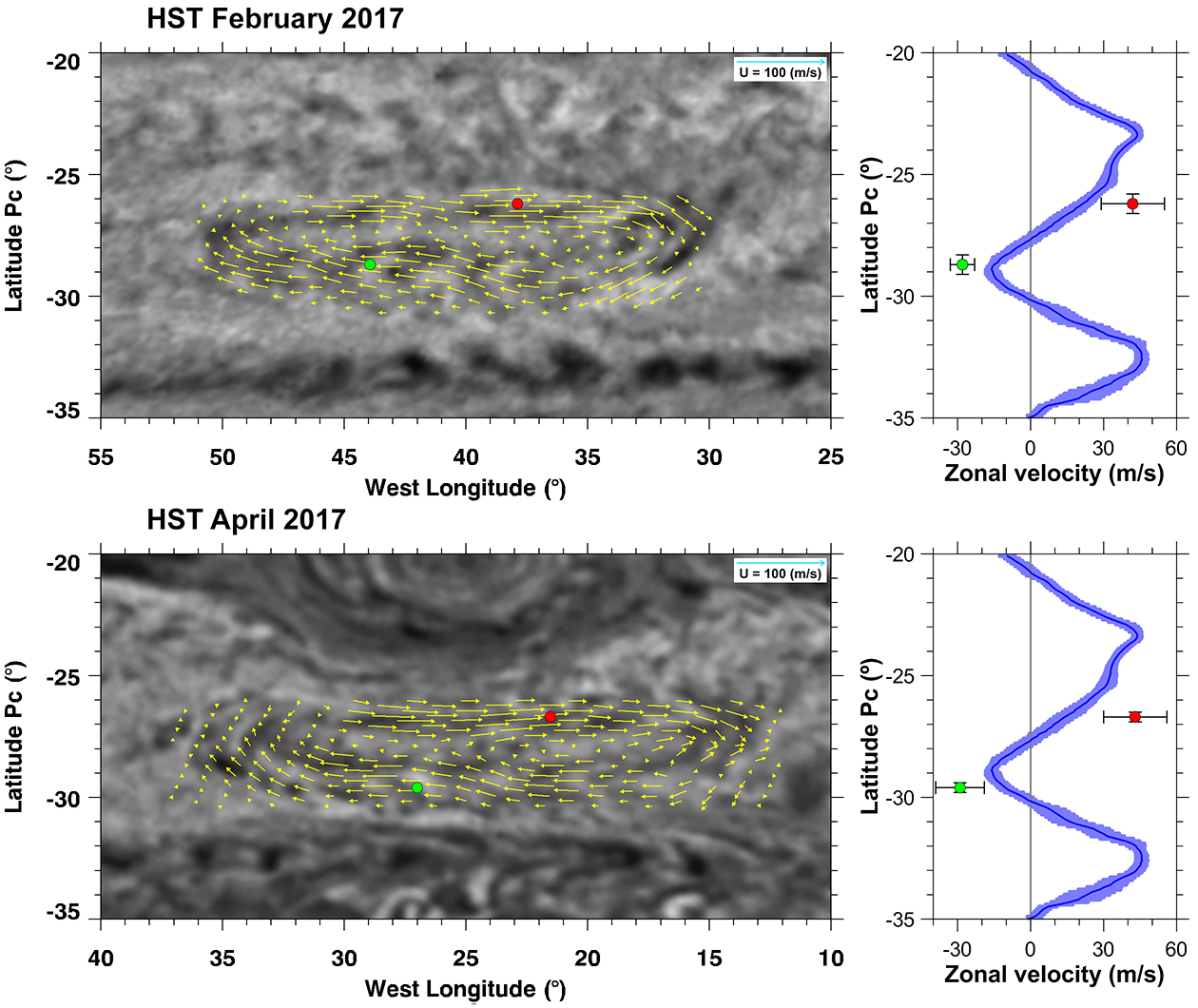}
\caption{Wind field in the STB Ghost retrieved from HST images taken on 1-2 February 2017 (top) and on 3 April 2017 (bottom) with images in both cases separated by about $10\,$hrs. The filter used in both cases is the F631N. The original measurements have been interpolated in a continuous two-dimensional field and are shown here at intervals of about $0.8^{\circ}$. The zonal-wind profile measured by \citet{Hueso_Jupiter_before_Juno_2017} is shown at the right side with a red and a green point indicating typical average velocities in the outer collar of the cyclone.}
\label{figure HST bidimensional}
\end{figure}

We also measured the wind field in the STB Ghost with HST images acquired in February and April 2017. Navigation errors were smaller than a pixel and the time separation between images allowed for more precise measurements. We used the PICV image-correlation software to obtain the wind field in the Ghost. The measurements were done using square correlation boxes with sides of $1$-$3^{\circ}$ and visually checking the identifications proposed by the algorithm and the correlation maps for individual details. Figure \ref{figure HST bidimensional} shows the results of these measurements in both dates compared with values of the zonal winds measured over HST images acquired in February 2016 \citep{Hueso_Jupiter_before_Juno_2017}. These HST images were also analysed by \citet{Tollefson_2017} who retrieved meridional profiles of zonal winds that are nearly identical to those shown in this figure (differences of less than $4\,$ms$^{-1}$ over this domain).

In the February 2017 images we obtained for the STB Ghost motions peak zonal wind speeds of ($52 \pm 10$)$\,$ms$^{-1}$ in its north limit and ($-39 \pm 10$)$\,$ms$^{-1}$ in its south limit. In the April 2017 images we obtained maximum wind speeds of ($64 \pm 10$)$\,$ms$^{-1}$ and ($-49 \pm 10$)$\,$ms$^{-1}$ for the outer northern and southern edges, respectively, consistent with the February measurements within the uncertainties. Visual tracking of a small number of selected features also resulted in similar values of the winds. Independent measurements of the wind speeds in the Ghost in February 2017 are given by \citet{Rogers_2019}, who reports peak zonal wind speeds of ($58 \pm 10$)$\,$ms$^{-1}$ in the northern limit and ($-35 \pm 10$)$\,$ms$^{-1}$ in the southern limit. These numbers are very similar to the wind field in the elongated cyclonic barges observed by the Voyager spacecraft at the NEB quoted above \citep{Hatzes_Voyager_Barge_1981}.

Due to the difficulties related to the short time differences in the measurements obtained from the JunoCam images, and since HST results in both dates are consistent, we favour the global wind speeds obtained from these HST images. However, HST images cannot resolve possibly faster motions in the outermost slightly red ring of material circulating the Ghost.

In February 2017, the STB Ghost was located at a planetocentric latitude of $\varphi_{pc}=(-28 \pm 0.5)^{\circ}$ with a size of $(23,000 \pm 600)\,$km x $(4,800 \pm 600)\,$km giving a maximum relative vorticity of $\zeta_{max}=\frac{V_{T_{max}} L}{\pi a b}=(-2.9 \pm 0.9) \cdot 10^{-5}\,$s$^{-1}$ where $V_{T_{max}}=52\,$ms$^{-1}$ is the maximum tangential velocity, $L$ is the perimeter of the ellipse of the vortex and $a$ and $b$ are the semi-major axes. From the zonal-wind profile from 2016 HST observations, the relative ambient vorticity at this latitude is $\zeta_{rel}=-1.28 \cdot 10^{-5}\,$s$^{-1}$, which is comparable to the planetary vorticity at this latitude, $f=-1.65 \cdot 10^{-5}\,$s$^{-1}$. Thus, the relative vorticity of the STB Ghost was approximately $2.3$ times the ambient relative vorticity.

\begin{figure}[h]
	\centering
		\includegraphics[angle=0, width=1.0\textwidth]{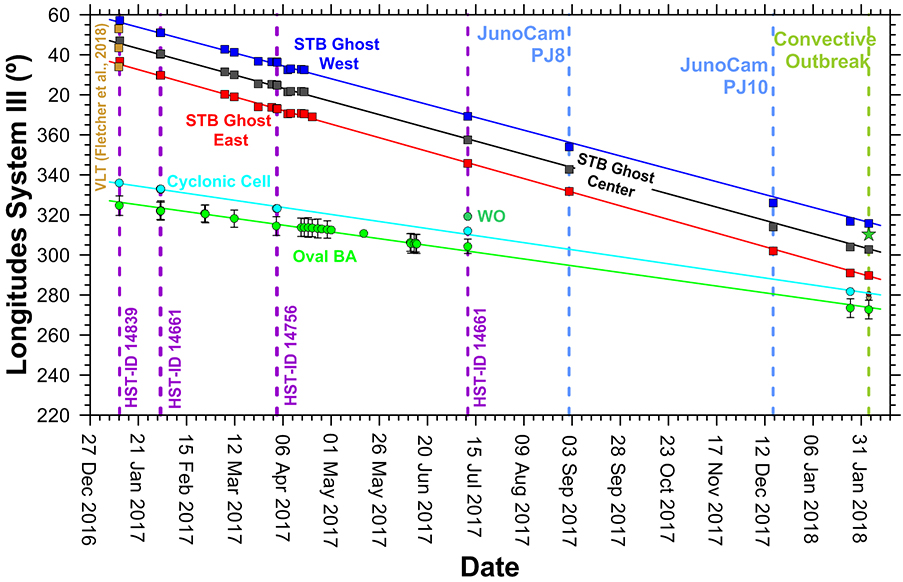}
	\caption{Longitudinal position over time of the STB Ghost, Oval BA, a cyclonic cell and a white oval (WO) west of Oval BA from amateur, HST, VLT and JunoCam images. Circles correspond to Oval BA (green filled circles), the cyclonic cell to its northwest (cyan filled circles) and WO (light green circle) to the southwest of Oval BA. Squares represent the STB Ghost central longitude (grey filled squares), its west limit (blue squares) and its east limit (red squares). The star on the right part of the plot represents the outbreak of the first storm. Linear fits to each set of measurements are also shown. Dates of high-resolution observations are indicated in the figure.}
    \label{figure_tracking_ghost}
\end{figure}

We also examined the drift rate of the STB Ghost and other major features in the STB to characterize the region. Figure \ref{figure_tracking_ghost} shows longitude positions of the STB Ghost, Oval BA and other atmospheric systems in the STB from December 2016 to February 2018. The planetocentric latitudinal position of the STB Ghost was $(-28.1 \pm 0.6)^{\circ}$ and the anticyclone Oval BA was located at latitude $(-29.6 \pm 0.2)^{\circ}$. Oval BA was accompanied by a cyclonic cell at its northwest side at $(-27.7 \pm 0.3)^{\circ}$ and a white oval was located southwest of Oval BA at latitude $(-29.8 \pm 0.5)^{\circ}$. Measurements over HST and amateur images show that the Ghost gradually increased in size from $(22,500 \pm 600)\,$km $\times (5,300 \pm 600)\,$km on 11 January 2017 to $(25,700 \pm 600)\,$km $\times (5,500 \pm 600)\,$km on 11 July 2017 until it reached a size of $(28,000\pm1,400)\,$km $\times (5,500 \pm 1,200)\,$km the day of the onset of the convective storm (with this last measurement based on amateur images). During 2017 the STB Ghost was steadily approaching Oval BA. Both vortices were separated by about $(17 \pm 2)^\circ$ the day of the beginning of the convective storm. Linear fits to the longitudinal positions of Oval BA and the STB Ghost resulted in longitudinal drift rates of $(-0.135 \pm 0.002)\,^\circ$day$^{-1}=(1.67 \pm 0.02)\,$ms$^{-1}$ and $(-0.268 \pm 0.002)\,^\circ/$day$^{-1}=(3.36 \pm 0.03)\,$ms$^{-1}$, respectively. At the latitude of the Ghost, the mean zonal-wind speed is $\bar{u}=(-7 \pm 3)\,$ms$^{-1}$ according to the 2016 HST zonal wind profile. The Ghost was drifting at a speed of about $u_{G}=(3.36 \pm 0.03)\,$ms$^{-1}$. Thus, the Ghost moved eastward relative to the mean background velocity $\bar{u}$ with a drift speed difference of  $u_{G}-\bar{u}=(10 \pm 3)\,$ms$^{-1}$.

\subsection{The convective eruption}
\label{section characterization storms}

\begin{figure}[htbp]
\centering
\includegraphics[angle=0, width=0.70\textwidth]{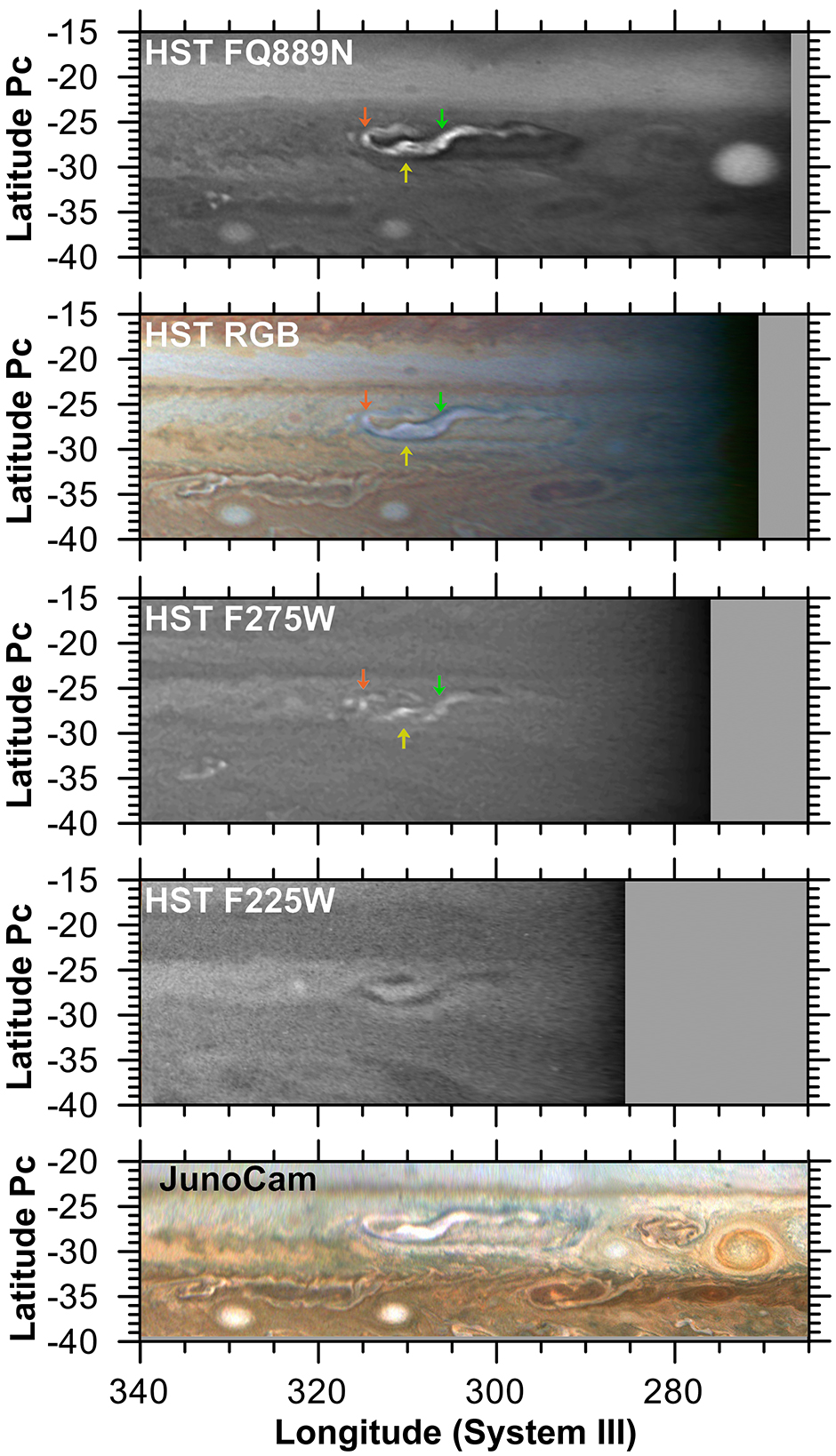}
\caption{High-resolution HST and JunoCam observations on 7 February 2018. All images have been processed to remove limb-darkening effects and with high-pass filters to better show the small-scale details. The locations of the three storms are marked with arrows in the HST images. Green represents the first storm, orange-red the second one, and yellow the third storm. The HST colour composition image was made assigning the filters F631N, F502N and F395N to the RGB channels. All the images were acquired with a time difference of less than 3 hours.}
\label{figure HST beginning storm}
\end{figure}

Amateur astronomer Anthony Wesley reported on 4 February 2018 the presence of a small but bright spot inside the STB Ghost near its west edge. The same spot had been observed one Jupiter rotation earlier by four other amateur astronomers, including particularly high-resolution observations by Damian Peach. Previous observations on 3 February 2018 did not show comparable bright spots. Figure \ref{figure onset storms} shows the first observations of the storm, compared with one rotation earlier. The convective nature of this feature was confirmed on images obtained with a filter at the strong methane absorption band at 890 nm, indicative of high clouds. Later images showed that the storm evolved into an elongated S shape with a northern branch drifting to the east and a southern branch to the west, which is reminiscent of the shape acquired by convective outbreaks in the cyclonic SEB \citep{Sanchez-Lavega_Icarus_1996, Hueso_JGR_2002, Fletcher_2017_SEB_storms}. Two days later, on 6 February 2018, a second convective spot appeared to the west of the first one (Figure \ref{figure onset storms}). On 7 February 2018, HST and JunoCam observations (Figure \ref{figure HST beginning storm}) showed that the storm system had evolved considerably, and signatures of a possible third convective nucleus could be noticed almost at the same location where the first storm erupted. 

\begin{figure}[htbp]
\centering
\includegraphics[angle=0, width=0.50\textwidth]{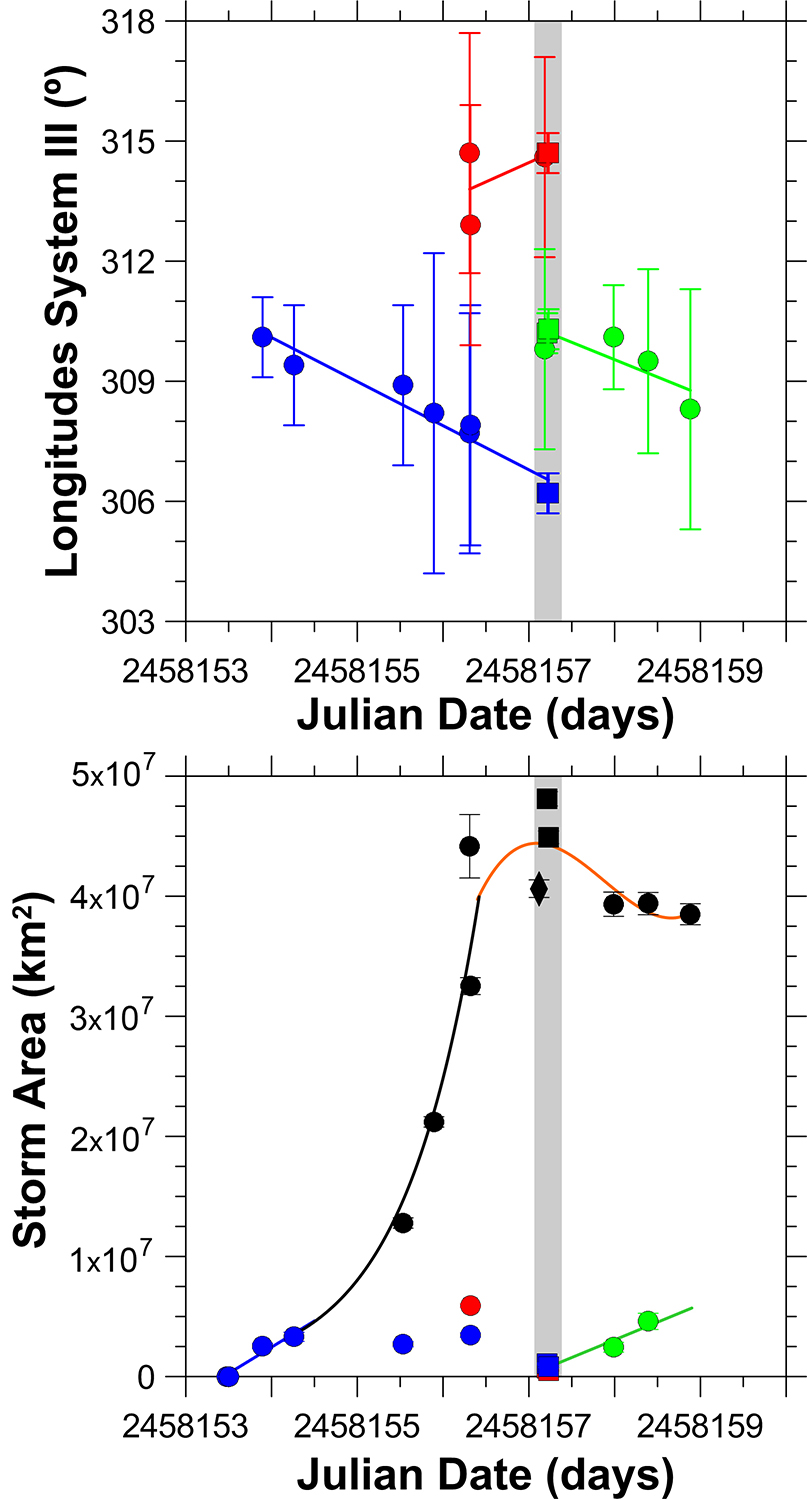}
\caption{Evolution of the storms over time. Top graph: longitudinal position of the bright spots as a function of time showing the possible presence of three different convective storms activated on different dates (in blue the first, in red the second and in green the third). Circles represent measurements on ground-based images, squares measurements on HST images. The light grey shaded region shows the period covered by HST images. A linear fit to each dataset is also shown. Bottom graph: Evolution of the bright cloud area as a function of time. Black filled circles indicate the total area covered by the storms' bright clouds and the light grey box represents measurements from HST observations. The diamond represents the total area of the storms measured on JunoCam images. The blue and green lines are linear fits (corresponding to the first and third storm respectively), the black line is an exponential fit and the orange one is a polynomial fit of 3\textsuperscript{rd} degree.}
\label{figure STB storms tracking}
\end{figure}

We tracked the positions of these convective sites and their sizes as a function of time. Figure \ref{figure STB storms tracking} shows these measurements. The characteristics of these storms are the following: 

\begin{itemize}
\item The first storm erupted at longitude $(310.1 \pm 1.0)^{\circ}$ and latitude $(-27.9 \pm 1.0)^{\circ}$. 
The size of the bright spot was $(1,900\pm1,000)\,$km $\times (1,600\pm1,000)\,$km and it was active over $\sim$4--5 days, moving with a drift rate of $(-1.1\pm0.1)\,^{\circ}/$day$^{-1}$ and an eastward velocity of $u=(14\pm1)\,$ms$^{-1}$. This is approximately $4$ times the drift speed of the Ghost and in the same eastward direction. If we compare the location and drift rate of the storm with the internal wind field previously obtained (see Figure \ref{figure HST bidimensional}) this velocity is opposite to that of the internal wind field for that location, which was around $u=-10\pm15\,$ms$^{-1}$. According to the HST zonal wind profile, the mean zonal-wind speed at the latitude of the storm is $\bar{u}=-4 \pm 4\,$ms$^{-1}$. The storm drifted in the opposite direction with a velocity $3.5$ times higher.

\item The second storm  erupted at longitude $(314.7 \pm 3.0)^{\circ}$ and latitude $(-26.2 \pm 2.5)^{\circ}$. The size of the bright spot was $(2,700\pm2,000)\,$km $\times (2,500\pm2,000)\,$km, and it was active over $\sim$1--2 days, moving with a drift rate of $(+1.0\pm0.8)^{\circ}/$day and a westward velocity of $u=(12\pm10)\,$ms$^{-1}$. This is about $3.6$ times the drift speed of the Ghost and in the opposite direction (westward). The motion of this storm cannot be compared with the internal motions of the Ghost or to the zonal-wind profile due to the large uncertainty in the latitude of the storm.

\item The third storm erupted at longitude $(310.2 \pm 0.5)^{\circ}$ and latitude $(-27.5 \pm 0.5)^{\circ}$. The size of the bright spot was $(1,500 \pm 600)\,$km $\times (900 \pm 600)\,$km and it was active over $\sim$2-3 days, moving with a drift rate of $(-0.9 \pm 0.3)\,^{\circ}/$day$^{-1}$ and an eastward velocity of $u=(11 \pm 4)\,$ms$^{-1}$, which is about $3.3$ times the drift speed of the Ghost and in the same eastward direction. The internal wind field of the Ghost at the location of this storm was $u=(-2 \pm 14)\,$ms$^{-1}$. According to the HST zonal-wind profile, the mean zonal-wind speed at the latitude were the third storm was observed is $\bar{u}=3 \pm 4\,$ms$^{-1}$. The storm drifted in the same direction with a velocity $3.7$ times higher.
\end{itemize}

The differences in the drift rates of the three storms and the internal wind field are probably related to the intense meridional shear of the zonal component of the internal wind field in the Ghost, but could also be a signature of different motions inside the cyclone at lower levels if the storms are deeply rooted \citep{Sanchez-Lavega_Nature_NTBD_2008}.

The second plot in Figure \ref{figure STB storms tracking} shows the approximate evolution of the area covered by each individual convective storm and the bright features spreading from them and forming elongated tails. At the beginning of the convective activity the nucleus of the first storm quickly grew linearly until reaching a stable size and later decreasing. The size evolution of the second and third storms were more difficult to evaluate, but the third storm behaved almost identical to the first one. The total area of the bright material spreading from the storms underwent a linear growth phase, followed by a rapid exponential growth until reaching a maximum size followed by a slow decrease. 

Through mass continuity, it is possible to relate the observed change rate of the area of the storms with the minimum vertical velocities at the cloud tops required to explain the observed divergence \citep{Hunt_Nature_1982_storms_growth, Hueso_Icarus_1998_Hot_spots}:

\begin{equation}
    w \sim H \, \nabla \cdot \vec{V}= H\, \frac{1}{A}\frac{dA}{dt},
\end{equation}
where $w$ is the minimum vertical velocity expected to cause the area $A$ of the clouds to grow at the observed rate and $H$ is the scale height at the cloud tops. This expression is based on a scale analysis argument that assumes that the vertical motion in the layer of outflow takes place over less than or equal to a scale height $H$. If we assume cloud tops at around $200\,$mbar, then $H=18\,$km. This assumption on the cloud tops and $H$ comes from the high brightness of the features in the methane absorption band image and the qualitative reflectivity of the clouds at different wavelengths observed in Figure \ref{figure HST beginning storm}. Within this approximation, the vertical velocity at the cloud tops would be between $0.3$ and $3\,$m$/$s. These values are consistent with those of typical jovian storms \citep{Hunt_Nature_1982_storms_growth, Hueso_Icarus_1998_Hot_spots}. However, note that this expression only places a minimum limit to the vertical updrafts and that updrafts that can be much narrower than the observed expanding cloud tops could produce vertical velocities that can be one or two order of magnitude larger \citep{Hueso_JGR_2002}.

\subsection{Evolution of the STB Ghost after the eruption}

\begin{figure}[htbp]
\centering
\includegraphics[angle=0, width=1.0\textwidth]{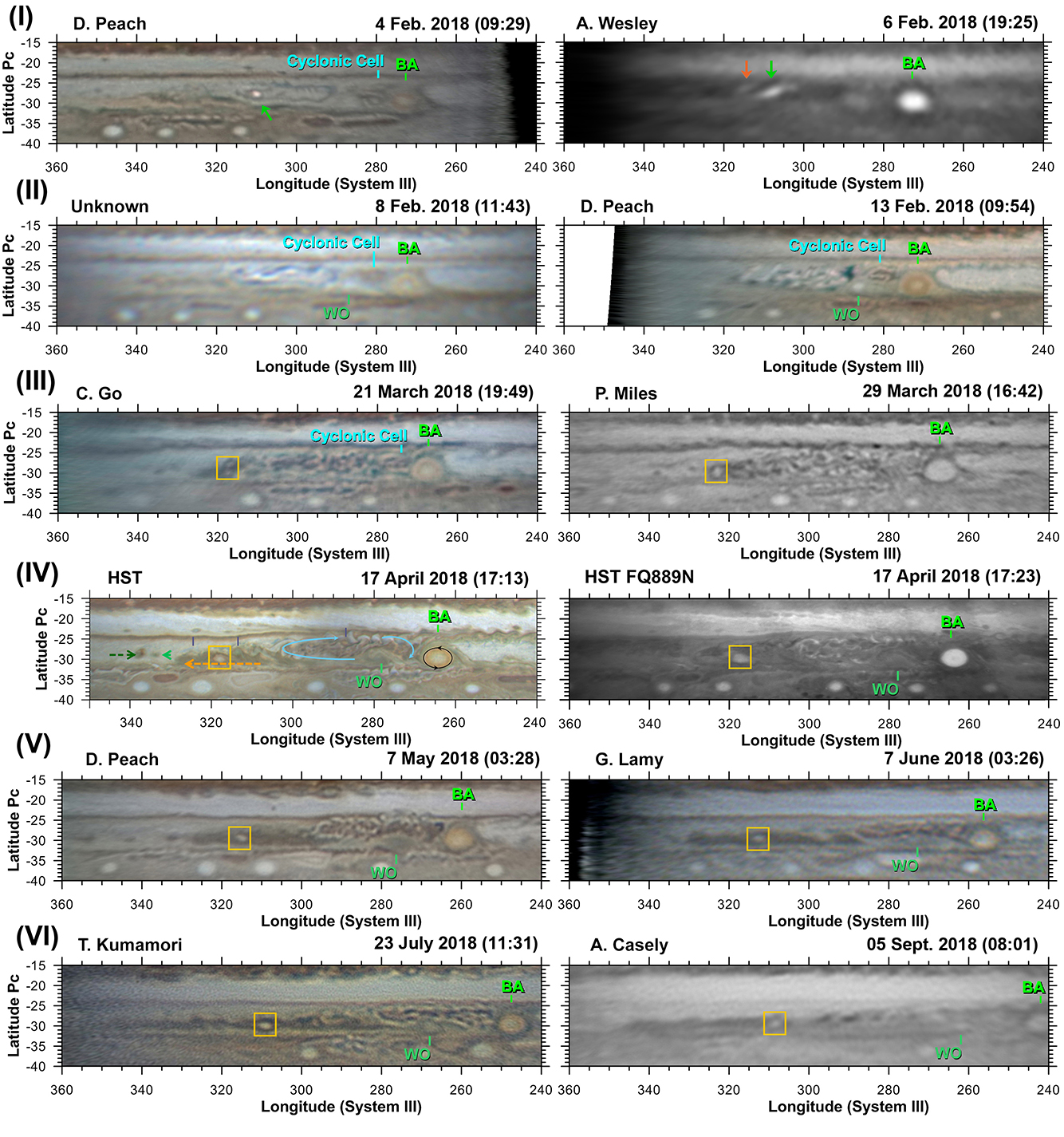}
\caption{Cylindrical maps of the perturbed region showing the different phases of the evolution of Jupiter's 2018 South Temperate Belt Disturbance: (I): Outbreak of the storms,  (II): bright filaments and dark features, (III): full disturbance, (IV): generation of a dark tail southwest of the STB Ghost, (V): extended evolution, and (VI): apparent dissipation. Date and time of the observations and name of the observer of each image are indicated. The positions of Oval BA, the cyclonic cell to its west and a white oval WO are highlighted. Also the outbreaks of the first two storms are indicated by arrows (the first storm in green and the second one in a orange-red colour) and two ovals that merged are highlighted with a dark-yellow box. Other features are highlighted with arrows. The HST colour  map was built from images in F631N, F502N and F395N filters.}
\label{figure STBD evolution 1-2-3-4-5-6}
\end{figure}

The interaction of the storm material with the Ghost circulation resulted in the development of significant turbulence initially confined to the interior of the Ghost. Turbulent patterns evolved over months without clear signatures of further convective storms. However, small bright patches of clouds were frequent in the north side of the Ghost in very high-resolution methane-band images, such as those obtained by HST and the PlanetCam UPV/EHU instrument (Figure \ref{figure PlanetCam}). We consider six different phases for the evolution of this event that are illustrated sequentially in Figure \ref{figure STBD evolution 1-2-3-4-5-6}.

\begin{itemize}
    \item I: Convective outbreaks (from 4 to 7 Feb. 2018). Figure \ref{figure HST beginning storm} shows high-resolution observations of the most active part of this phase. Figure \ref{figure STBD evolution 1-2-3-4-5-6} shows amateur observations detailing the initial evolution of the outbreaks including observations in the methane absorption band.
    
    \item II: Bright filaments and dark features (from 7 to 15 Feb. 2018). A series of bright filaments and large dark features circulated in the STB Ghost with some of the black features acquiring relatively large sizes. A white oval at $29.8^{\circ}$S started to interact with the south branch of the Ghost.
    
    \item III: Disturbed Ghost (from 15 Feb. 2018 to late March 2018). The turbulent patterns that were formed at the end of the previous phase were confined inside the Ghost area. Some small dark features (possibly anticyclones because of their latitudinal position) were expelled from the west side of the Ghost moving westward (examples are shown with green arrows in the HST panel in Figure \ref{figure STBD evolution 1-2-3-4-5-6}). Additional ovals were observed inside the disturbed region southwest of the Ghost, with two of them merging and forming a new large oval (dark-yellow box in Figure \ref{figure STBD evolution 1-2-3-4-5-6}). This new oval has been present since its formation in March 2018 until the last image analysed in this work that was obtained in February 2019. During this period it became evident that the rate at which the Ghost approached Oval BA had largely diminished by almost 50\% with a drift rate of $(-0.170 \pm 0.008)^\circ$day$^{-1}$. This slower drift rate could have started at the time of the convective storms but it is not possible to assess this quantitatively as a few days are needed to capture a precise measurement of the drift rate. This reduction of the drift rate was accompanied by a clear expansion towards the west. The close approach to the Oval BA resulted in a strong interaction with the cyclonic region northwest of Oval BA, which started at the beginning of March 2018 and lasted for a couple of months. This interaction was accompanied by a simultaneous interaction with the white oval southwest of Oval BA that curved the southeast part of the Ghost northwards and lasted at least until August 2018. We note that as a consequence of this interaction, the external ring of the long-lived anticyclone Oval BA became darker and less pronounced. However the drift rate of Oval BA did not seem to be perturbed.

    \item IV: Generation of a dark tail and zonal expansion (from late March 2018 to early May 2018). Around 1 April 2018 a very large dark structure on the southwest side of the STB Ghost became evident. This structure extended progressively to the west forming an elongated dark tail. Panel (IV) in Figure \ref{figure STBD evolution 1-2-3-4-5-6} shows maps obtained from HST observations of the area with the interaction with Oval BA and the morphology of the elongated system. We highlight in the colour HST map the circulation of the Ghost and Oval BA and the drift directions of selected features like the large dark tail created. The HST image in the methane absorption band shows some bright features inside the Ghost, probably indicating a certain degree of vertical motions. The Ghost went over the white oval in its south flank at $29.8^{\circ}$S forcing the Ghost to bend around it. This anticyclone has been observed almost continuously at least since January 2018, except during March 2018, when it was probably obscured by the turbulence in the remains of the Ghost. 

We used PICV to measure the wind field in the HST images obtained on 17 April 2018 (Figure \ref{figure_STBD_bidimensional}). The dark streak region coming from the STBD in its west side shows typical motions of the local zonal jets without apparent perturbations. The cyclonic circulation of the Ghost is limited longitudinally from $303^{\circ}$ to $270^{\circ}$, which is slightly longer than the size of the Ghost before the perturbation, and interacts strongly with the anticyclonic white oval southwest of Oval BA. We obtained a maximum wind speed of $(65 \pm 10)\,$ms$^{-1}$ in the northern edge and of $(-54 \pm 10)\,$ms$^{-1}$ in the southern edge, which is comparable to the wind values obtained before the convective perturbation and almost identical to the measurements in April 2017. Independent measurements of the wind speeds in the Ghost in February 2018 are given by \citet{Rogers_2019} and are also in agreement within the measurement uncertainties. Wind motions inside Oval BA cannot be retrieved with confidence due to the large angular rotation of features inside the oval in the time interval of $10\,$hr between the images used.

\begin{figure}[h]
\centering
\includegraphics[angle=0, width=1.0\textwidth]{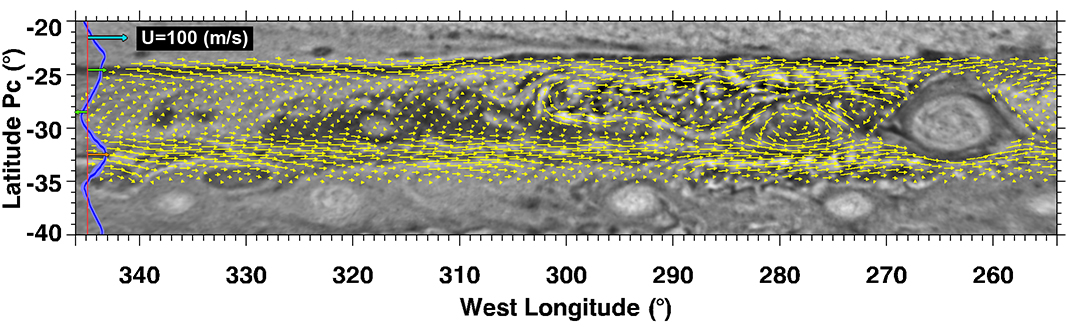}
\caption{Wind field on 17 April 2018 from HST images. Maximum wind speeds of $(65 \pm 10)\,$ms$^{-1}$ in the northern edge and of $(-54 \pm 10)\,$ms$^{-1}$ in the southern edge were measured. The measurements have been interpolated into a regular grid and vectors are drawn at intervals of $\sim 0.8^{\circ}$. Wind vectors in Oval BA are not drawn. The high angular rotation of Oval BA, and the large time interval between the observations used made obtaining wind motions inside it unreliable. The zonal-wind profile measured by \citet{Hueso_Jupiter_before_Juno_2017} is also shown at the left side of the map with a blue solid line. The zero-velocity level is shown with a red line and a blue reference wind vector shows the scale for the wind vectors and the zonal wind profile. Green arrows show the intensity and direction of mean winds in the north and south edges of the remains of the Ghost. An orange arrow shows winds in the South limit of the perturbed region by the interaction of the Ghost and the WO and is similar to the zonal winds.}
\label{figure_STBD_bidimensional}
\end{figure}

    \item V: Extended evolution (from early May 2018 to late June 2018). The southwest structure created in April continued its westward drift and zonal expansion making obvious the separation between the Ghost and the tail. New anticyclones were expelled to the west. The STB Ghost continued to be located west of Oval BA as well as the anticyclone on its southeast side. By the end of May, a medium-size structure separated from the southwest structure's east side. The PlanetCam UPV/EHU methane-band observations in May 2018 (see figure \ref{figure PlanetCam}) showed bright and dark structures inside the Ghost pointing to possible vertical motions still present even at that late date. The white oval at the south flank of the Ghost drifted westward with respect to the Ghost partially shaping its structure.

\item VI: Dissipation and possible reactivation (after June 2018). The southwest structure continued its zonal elongation with occasional features separating from it. Finally, in September 2018 a large section of the east side of the southwest structure tore apart, generating five small dark features. The JunoCam instrument observed this area on Juno's 16\textsuperscript{th} perijove on 29 October 2018 with Jupiter near solar conjunction. The Ghost had largely dissipated and apparently merged with the cyclonic cell northwest of Oval BA (Figure \ref{figure STBD JunoCam PJ16}). The area was observed by JunoCam again on the next perijove on 21 December 2018. The new images showed a morphology of the cyclonic cell slightly enlarged and with strong turbulence. The morphology of the cyclonic cell evoked to some extent the excited phases of the STBD, with strong turbulence inside the cyclone. JunoCam observations in December 2018 showed that this cyclonic cell expanded to a size of 20$^{\circ}$ of longitude, but the long dark tail formed from the turbulent Ghost was no longer observed. During January and February 2019 the cyclonic cell kept its longitudinal size between $15^{\circ}$ and $20^{\circ}$, and a morphology similar to the one observed by JunoCam in December 2018 (bottom panel in Figure \ref{figure STBD JunoCam PJ16}) or to ground-based observations of the active Ghost. Thus, the turbulence inside the Ghost-cyclonic cell system seemed to reactivate between October and December but there is no evidence of active convection in this period. Later observations over 2019 showed a continuation of this turbulence without signatures of active convection.
\end{itemize}

\begin{figure}[htbp]
\centering
\includegraphics[angle=0, width=1.0\textwidth]{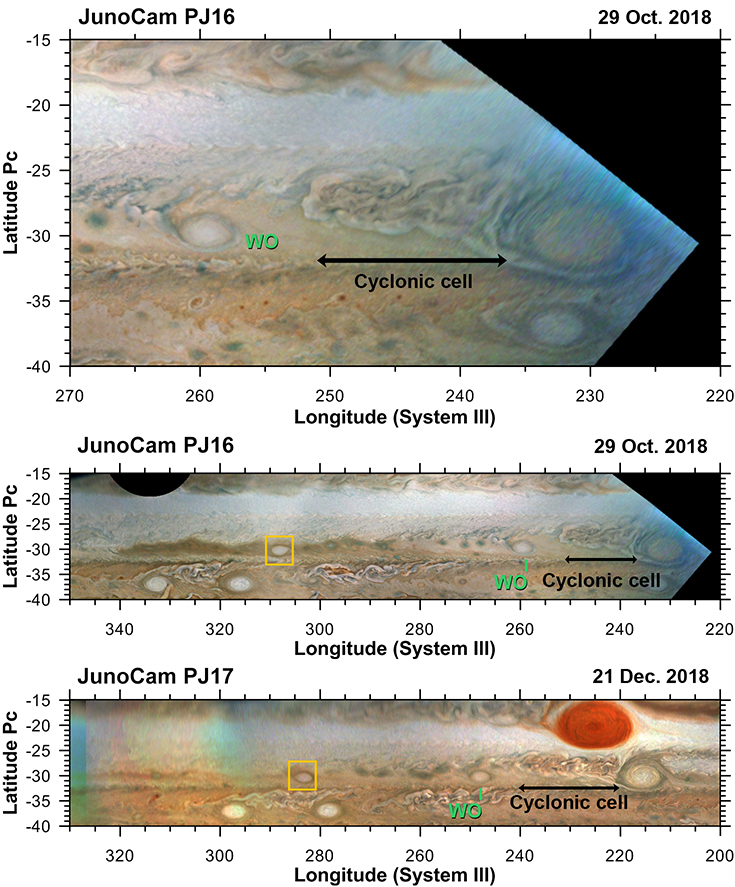}
\caption{Maps of the STB from JunoCam images on 29 October 2018 (top at high-resolution and middle image with a broader context) and on 21 December 2018 (bottom). The position of the white oval shown on Figures \ref{figure HST Ghost before storms} and \ref{figure STBD evolution 1-2-3-4-5-6} is highlighted and the dark-yellow box indicates the oval generated from the merger of two ovals as shown on Figure \ref{figure STBD evolution 1-2-3-4-5-6}. The cyclonic cell generated from the merger of the STB Ghost and the previous cyclonic cell northwest of Oval BA is also highlighted.}
\label{figure STBD JunoCam PJ16}
\end{figure}

\begin{figure}[htbp]
	\centering
		\includegraphics[angle=0, width=1.0\textwidth]{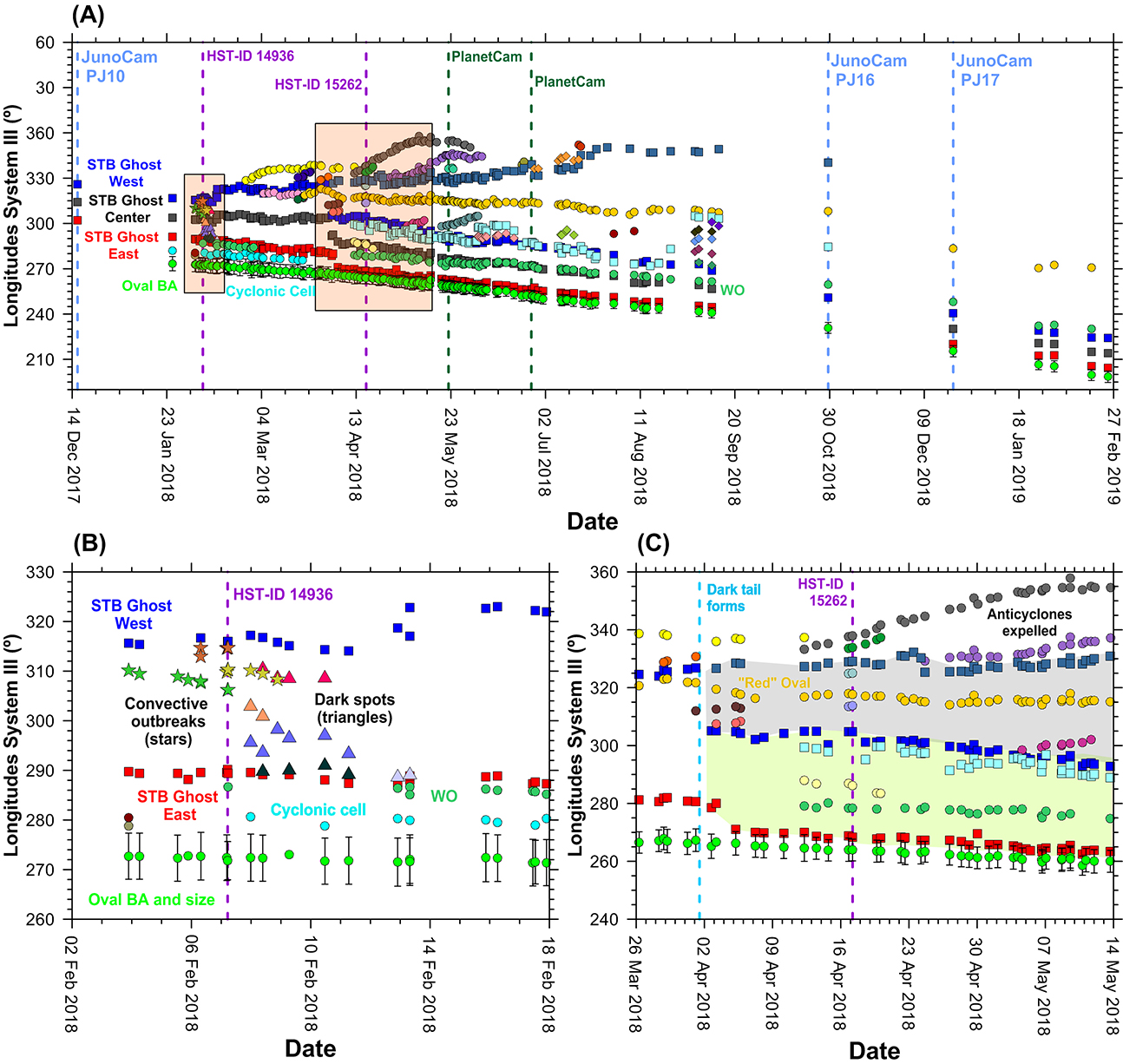}
	\caption{Features tracked in the 2018 Jupiter's South Temperate Belt Disturbance. Panel (A) shows the tracking of the features since the start of the convective activity and is a continuation of Figure \ref{figure_tracking_ghost}. Panel (B) covers the onset of the convective activity. Panel (C) shows details of the evolution of the different features shortly after the formation of the large dark tail. On all panels the large filled green circles show Oval BA and its longitudinal extent. Other coloured circles represent the cyclonic cell at its northwest (cyan), the white oval at the southwest of Oval BA (light green), the long-lived oval generated in March (dark-yellow), and other smaller and shorter-lived ovals and anticyclones expelled from the Ghost. Squares show the east and west edges and the centre of the STB Ghost and the large dark tail. Stars indicate the convective outbreaks, triangles the dark spots circulating the Ghost and diamonds some other small structures. On panel (C) the light green and light grey shadow regions indicate the STB Ghost and large dark tail respectively. The dates when HST, JunoCam, PlanetCam or VLT data were available are also indicated.}
	\label{figure tracking STBD}
\end{figure}

In order to quantify the evolution of the STBD described above we tracked $45$ individual features since the date of the first storm until the amateur observations in February 2019 (Figure \ref{figure tracking STBD}). The features tracked are the STB Ghost, Oval BA, the individual convective outbreaks and the main ovals, anticyclones, cyclones, dark features, storms and large structures visible during the course of the disturbance. Some of these features were long-lived, surviving most of the overall activity, and others were short-lived with lifetimes of days. Panel (A) displays the timeline of the event. Panel (B) shows the features tracked during the convective stage and panel (C) shows the evolution of the system after the generation of the large dark tail including the westward motion of several dark spots expelled from the Ghost area and the generation of small short-lived ovals. The figure shows the apparent elongation of the Ghost, its approach to Oval BA, the merger of particular ovals and the tracking of the white oval (WO) southwest of Oval BA. The figure also shows how the southwest dark branch of this interaction elongated and broke again into smaller structures. Vortices highlighted in the previous figures are shown in these drift charts with the same colour codes.

Figure \ref{figure tracked features drift velocity} summarizes the information from these tracks showing the zonal drift speed of the features compared to the background zonal winds from HST images in February 2016 \citep{Hueso_Jupiter_before_Juno_2017}.  This zonal wind profile is nearly identical for the latitudes of interest to zonal wind profiles obtained also with HST data in February and December 2016 by \citet{Tollefson_2017}. A later analysis of zonal winds in Jupiter from HST data including data acquired in 2017 does not show temporal changes from 2016 and 2017 at the latitudes of the STB \cite{Johnson_PSS_2018}. The three storms moved with their own motions different from the background zonal-wind profile. Also, Oval BA, the STB Ghost and the oval formed from the merger of other two ovals on March 2018, and highlighted with a dark-yellow box on Figures \ref{figure STBD evolution 1-2-3-4-5-6} and \ref{figure STBD JunoCam PJ16}, drifted with a velocity different from the zonal-wind profile. The cyclonic cell northwest of Oval BA and the white oval drifted with the velocity of the zonal-wind profile at the latitudes where they were located.
 
\begin{figure}[h]
	\centering
		\includegraphics[angle=0, width=1.0\textwidth]{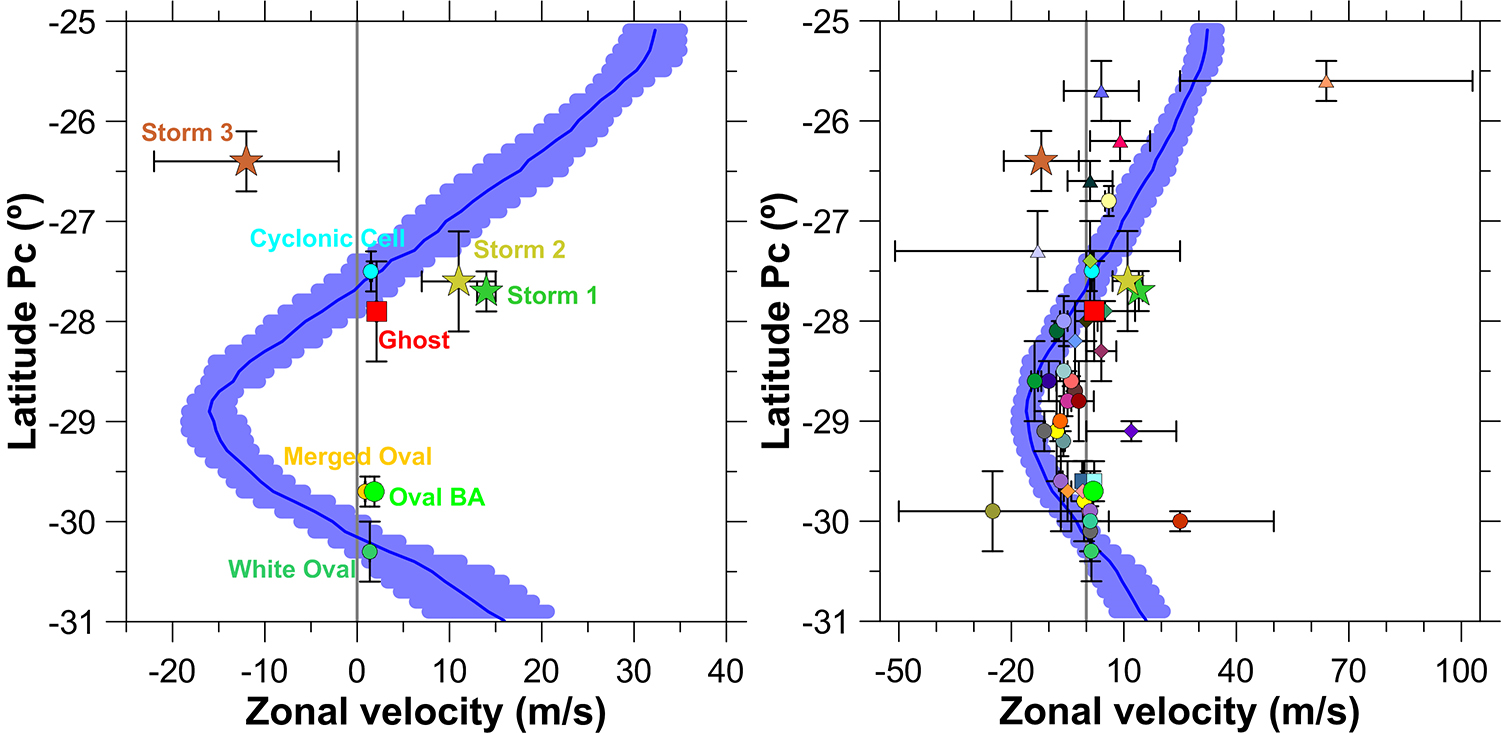}
	\caption{Zonal drift speed of the features tracked during the STBD. The left panel shows the zonal drift speed for the main features. The right panel the zonal drift speeds of all the features tracked. Symbols and colours are the same as in Figure \ref{figure tracking STBD}. Stars indicate the convective outbreaks, triangles the dark spots circulating the Ghost, circles represent ovals and features expelled westward from the STB Ghost and diamonds indicate other small structures. The red square indicates the STB Ghost, the cyan and green squares indicate the large dark tail. The blue line represents the HST 2016 zonal-wind profile from \citet{Hueso_Jupiter_before_Juno_2017}.}
	\label{figure tracked features drift velocity}
\end{figure} 
 
Most of the small features tracked (right panel in Figure \ref{figure tracked features drift velocity}) moved with velocities slightly different from the zonal-wind profile. These features generated inside the STB Ghost or in its long dark tail and moved at velocities intermediate from the nearly null drift rate of the STB Ghost and the westward zonal winds. The dark spots circulating the STB Ghost at the first stages of the disturbance (triangles on Figure \ref{figure tracked features drift velocity}) had velocities very different from the zonal-wind profile and moved following the internal motions in the STB Ghost. The two dots near $-30^{\circ}$ correspond to two small dark features expelled from the dark tail. The purple diamond near planetocentric latitude $-29^{\circ}$ corresponds to one of the features generated on the break-up of the long dark tail on September 2018 and moved differently to the zonal-wind profile. Table \ref{table drift rate features} in Appendix B details these measurements.

\subsection{Interaction with Oval BA}
The ensemble of measurements of the main features in the STB from 2017, 2018 and 2019 (Figures \ref{figure_tracking_ghost} and \ref{figure tracking STBD}) shows that there was a strong decrease of the drift rate of the STB Ghost from $(-0.268 \pm 0.002)^\circ$day$^{-1}$ to $(-0.170 \pm 0.008)^\circ$day$^{-1}$ from measurements before and right after the onset of the convective activity. However, Oval BA changed its drift rate only slightly as a consequence of its interaction with the STB disturbed Ghost and accelerated from a drift rate of $(-0.135 \pm 0.002)^\circ$day$^{-1}$ prior to the interaction with the STB Ghost to $(-0.152 \pm 0.001)^\circ$day$^{-1}$ after this interaction started. These changes were within the historical variations of the drift rate of Oval BA observed in previous years \citep{Garcia-Melendo_BA_I_2009}. The maps shown on Figure \ref{figure STBD evolution 1-2-3-4-5-6} also show that the colour of the external ring of Oval BA changed into a darker grey ring with faint green colour in HST images.

\subsection{Summary of the observational analysis}
A brief summary of the observed phenomenology is given below:

\begin{itemize}
    \item During 4 and 7 February 2018 three convective storms erupted inside an elongated cyclone known as the STB Ghost. At this time the STB Ghost was separated by $(17 \pm 2)^{\circ}$ from Oval BA. A small cyclonic cell and a white anticyclone were located between the STB Ghost and Oval BA.

    \item The storms generated strong turbulence inside the Ghost and formed dark features and bright filaments. Small dark features were expelled westward from the southwest limit of the STB Ghost. Two of these features merged and generated a long-lived oval. In April 2018 a large dark reddish tail was visible to the west of the STB Ghost. The STB Ghost interacted strongly with the cyclonic cell northwest of Oval BA until they merged. The Ghost also strongly interacted with the white anticyclone on its south side.

    \item The dark tail and the STB Ghost expanded zonally while they were separating from each other. Oval BA acted as a barrier to the eastward drift of the cyclonic system generated by the merger between the previous cyclonic cell and the STB Ghost.

    \item The dark tail dissipated after it broke down into several smaller dark features and completely  disappeared by December 2018. Between October and December 2018 the turbulence within the new cyclonic system seemed to reactivate. This occurred without clear signs of new convective storms.
\end{itemize}

\section{Numerical modelling} \label{section_EPIC}

We have used the Explicit Planetary Isentropic-Coordinate (EPIC) atmospheric model  (version 3.8) \citep{Dowling_1998} to simulate the phenomenology observed and gain insights into the nature of the convective activity and later evolution. 

EPIC is a General Circulation Model that solves the hydrostatic primitive equations using a finite-differences scheme and potential temperature, $\theta$, as the vertical coordinate. The model computes the evolution of potential vorticity, which is a conserved variable following the motion for inviscid and adiabatic flow and can be used as a tracer of the observed cloud patterns. A typical simulation with EPIC starts from a model atmosphere in geostrophic equilibrium where atmospheric perturbations are introduced to artificially initiate vortices or waves. While the model is not able to predict the onset of such systems, it is able to simulate in a realistic way their interaction with the zonal winds and other meteorological systems. This version of EPIC has been used extensively to model Jupiter's atmospheric vortices and their interactions \citep{Morales_BE_FA_merger_2003, Legarreta_EPIC_Jup_vortex_2008, Morales_EPIC_GRS_2013, Garcia-Melendo_BA_I_2009}.

Since EPIC uses isentropic coordinates it requires a vertically stable atmosphere and is not able to predict when convective disturbances will appear. Motions are adiabatic and vertical motions are only possible where there is heating. The heating can be introduced locally, simulating convective disturbances that violate the hydrostatic condition at their location, whereas remaining valid in the rest of the model domain. In these conditions EPIC is able to simulate the response of the atmosphere to convective disturbances whose energy is modulated by the user when running a simulation \citep{Garcia-Melendo_2005, Sanchez-Lavega_Nature_NTBD_2008, Sanchez-Lavega_NTBD_2017}. EPIC has also been used in other giant planets like Neptune \citep{LeBeau_Neptune_GDS_1998, Stratman_Neptune_EPIC_2001} and Saturn \citep{Sayanagi_Saturn_GWS_2007, del_Rio-Gaztelurrutia_Saturn_tripole_2018}, including studies of convective storms \citep{Sanchez-Lavega_2010_GWS_Nature_2011}. In studies of moist convective storms the user introduces heating sources that represent the size and intensity of the storms and that are initiated in agreement with the observations.

In this work we initialize a reference atmosphere where the STB Ghost, Oval BA and the convective storms are introduced sequentially on different days of the simulation and the model is used to observe their interaction. To avoid further complexities, we do not incorporate the cyclone and anticyclone northwest and southwest of Oval BA. After initiating each feature the model is allowed to evolve for some time until it stabilizes before introducing the next atmospheric system. EPIC is formulated using planetographic latitudes but we will show results from EPIC simulations in planetocentric latitudes to be consistent with the previous figures.

\subsection{Model atmosphere}

The model is initialized with a reference atmosphere in geostrophic balance defined in a longitudinal channel with periodic boundaries at its longitudinal limits. The reference atmosphere is defined by the vertical thermal profile, which determines the static stability, the zonal wind profile and the vertical wind shear. EPIC is initialized with a single thermal profile for the whole simulation domain and the three dimensional thermal structure is calculated later by imposing geostrophic equilibrium once the zonal winds are introduced.

\subsubsection{Model domain}

We used a domain of $160^{\circ}$ in longitude and $20^{\circ}$ in planetographic latitude ($18.75^{\circ}$ in planetocentric latitude), going from $-38.2^{\circ}$ planetocentric to $-19.5^{\circ}$, with a resolution of $0.16^{\circ}$ per grid point. We used a time step of $20\,$s that fulfils the numerical stability conditions required by the EPIC model, including those not directly linked to the Courant CFL condition, but with the hyperviscosity scheme used in EPIC \citep{Dowling_1998}. The vertical domain goes from $10\,$mbar to $7\,$bar and is divided in 8 layers, the last layer being located below the $7\,$bar pressure level. This bottom layer is an abyssal layer which is not allowed to evolve dynamically and represents the adiabatic interior of the planet. The top two layers are ``sponge" layers, where numerical dissipation is included to attenuate reflections of dynamical terms like gravity waves from the top of the model. A hyperviscosity term $\nu_{6}$ is also used to control numerical instabilities \citep{Dowling_1998}.

\subsubsection{Thermal structure}
\label{thermal_structure}

\begin{figure}[h]
	\centering
	\ContinuedFloat*
		\includegraphics[angle=0, width=1.0\textwidth]{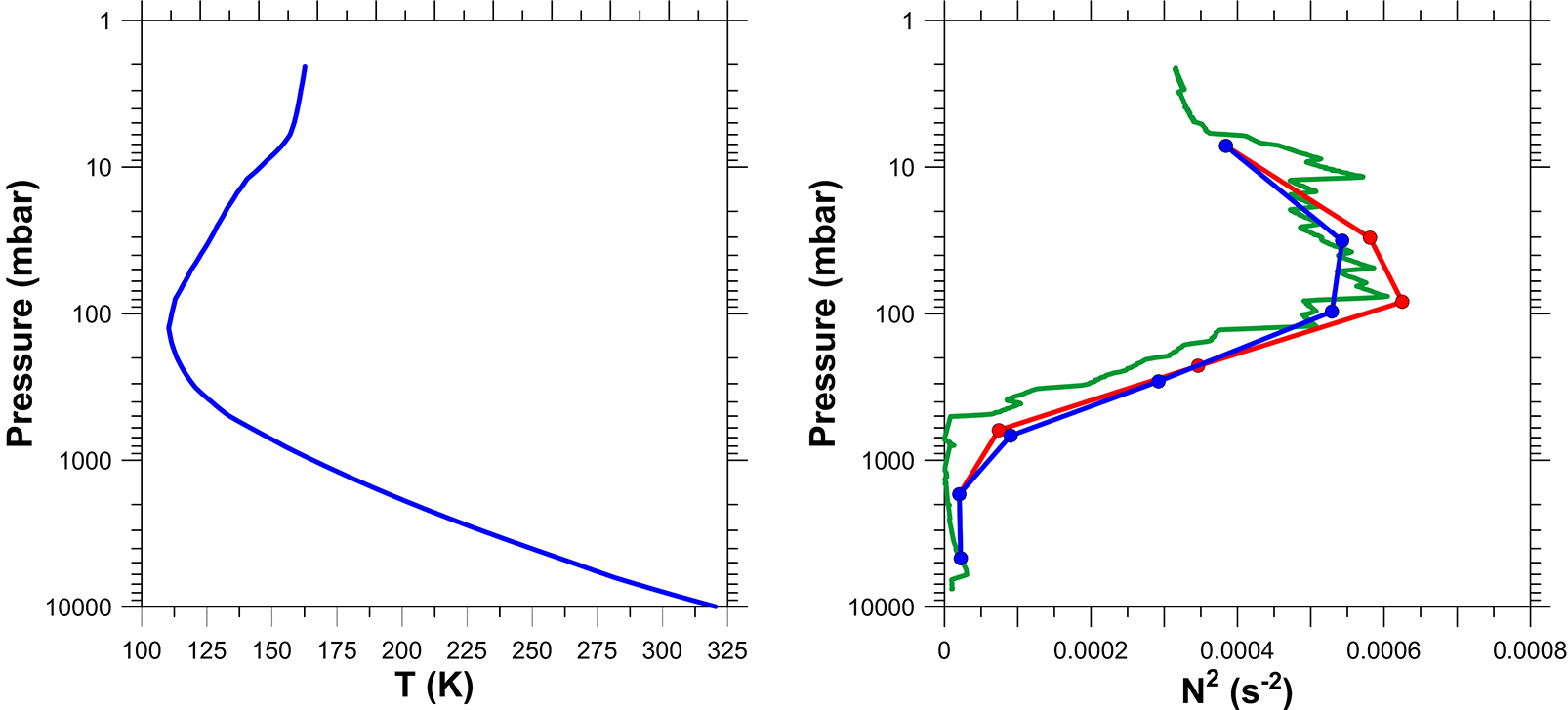}
	\caption{Thermal properties of the reference atmosphere. Left: Thermal profile from Voyager 1 and 2 radio-occultation experiment  \citep{Lindal_Voyager_radio_occultation_1981} above the visible clouds and extrapolated below following a moist adiabat as explained in the text. Right: Brunt--V\"{a}is\"{a}l\"{a} frequency. The green line is the squared Brunt--V\"{a}is\"{a}l\"{a} frequency computed from the thermal profile. EPIC computes values of the static stability associated to individual layers of the atmosphere that approximate but do not equal the continuous profile. Values used by EPIC are shown with lines and symbols in this plot. Because the model is in geostrophic balance, the presence of winds alters the static stability of the atmosphere. The blue line represents the profiles used in our nominal reference atmosphere with constant winds below the clouds and decaying winds above them. The red line represents the profile for a constant wind with no vertical shear.}
\label{figure EPIC atmosphere part1}
\end{figure}

The vertical thermal structure was based on an average of the thermal profiles obtained in the Voyager radio-occultation experiments at $-12^{\circ}$, the equator and $-60^{\circ}$ \citep{Lindal_Voyager_radio_occultation_1981}. Although the three profiles present substantial differences above the $100$-mbar level, they are very similar in the lower troposphere at around $700\,$mbar. This thermal profile could also differ from modern derivations of temperatures in Jupiter, and in particular in the STB, \citep{Fletcher_Icarus_2016} but it provides a first-order approximation to the jovian thermal structure that has been used successfully in many previous studies of Jupiter. This thermal profile was extrapolated below the $0.7$-bar level, assuming a wet adiabat with condensable species H$_2$O, NH$_3$ and NH$_4$SH as in previous works related to EPIC \citep{Garcia-Melendo_2005, Legarreta_EPIC_Jup_vortex_2008}. Figure \ref{figure EPIC atmosphere part1} shows the thermal profile used and the corresponding squared Brunt--V\"{a}is\"{a}l\"{a} frequency. The Brunt--V\"{a}is\"{a}l\"{a} frequency, $N=\sqrt{(g/\theta) (d\theta/dz)}$, where $g$ is the local acceleration of gravity, $\theta$ is the potential temperature, and z is the vertical coordinate, is a measure of the stability of the atmosphere with respect to vertical motions.

\subsubsection{Zonal Winds}
\label{zonal_winds}

We initialized the atmosphere with zonal winds coming from either (i) the Cassini flyby in December 2000 \citep{Porco_Cassini_Jup_winds_2003}, or (ii) zonal winds measured by \citet{Hueso_Jupiter_before_Juno_2017} using Jupiter images acquired by HST in 2016. The HST profile is essentially identical to zonal winds derived by \citet{Tollefson_2017} for the same dates. In both cases the reference zonal-wind profile corresponds to the visible ammonia cloud level located approximately at $680\,$mbar. Both the 2016 HST wind profile and the Cassini wind profile are very similar. However, differences between both profiles in the domain of the simulations are on the order of 5 ms$^{-1}$ and reach 12 ms$^{-1}$ at certain latitudes. Thus, simulations with the Cassini zonal wind profile could not match the drift rates of the observed atmospheric features and we considered the HST 2016 wind profile a better option. 

The vertical structure of the winds is introduced using a multiplicative normalization factor that varies as a function of altitude but is constant at all latitudes \citet{Legarreta_EPIC_Jup_vortex_2008}. The STB is a warm belt resulting in thermal winds that imply that winds decay with altitude above the cloud level \citep{Fletcher_Icarus_2016}. We tested winds decaying with altitude above the clouds until null values at $30\,$mbar and constant in depth below the clouds. This vertical structure of the zonal winds was considered as our ``nominal" case and is the same used on \citet{Legarreta_EPIC_Jup_vortex_2008}. We also tested constant winds in height and winds increasing in height. Figure \ref{figure EPIC atmosphere part2} summarizes the zonal winds and their vertical structure used in our simulations. Winds increasing in altitude with wind-shears stronger than the one shown in Figure \ref{figure EPIC atmosphere part2} produced unstable simulations.

\begin{figure}[h]
	\centering
	\ContinuedFloat
		\includegraphics[angle=0, width=1.0\textwidth]{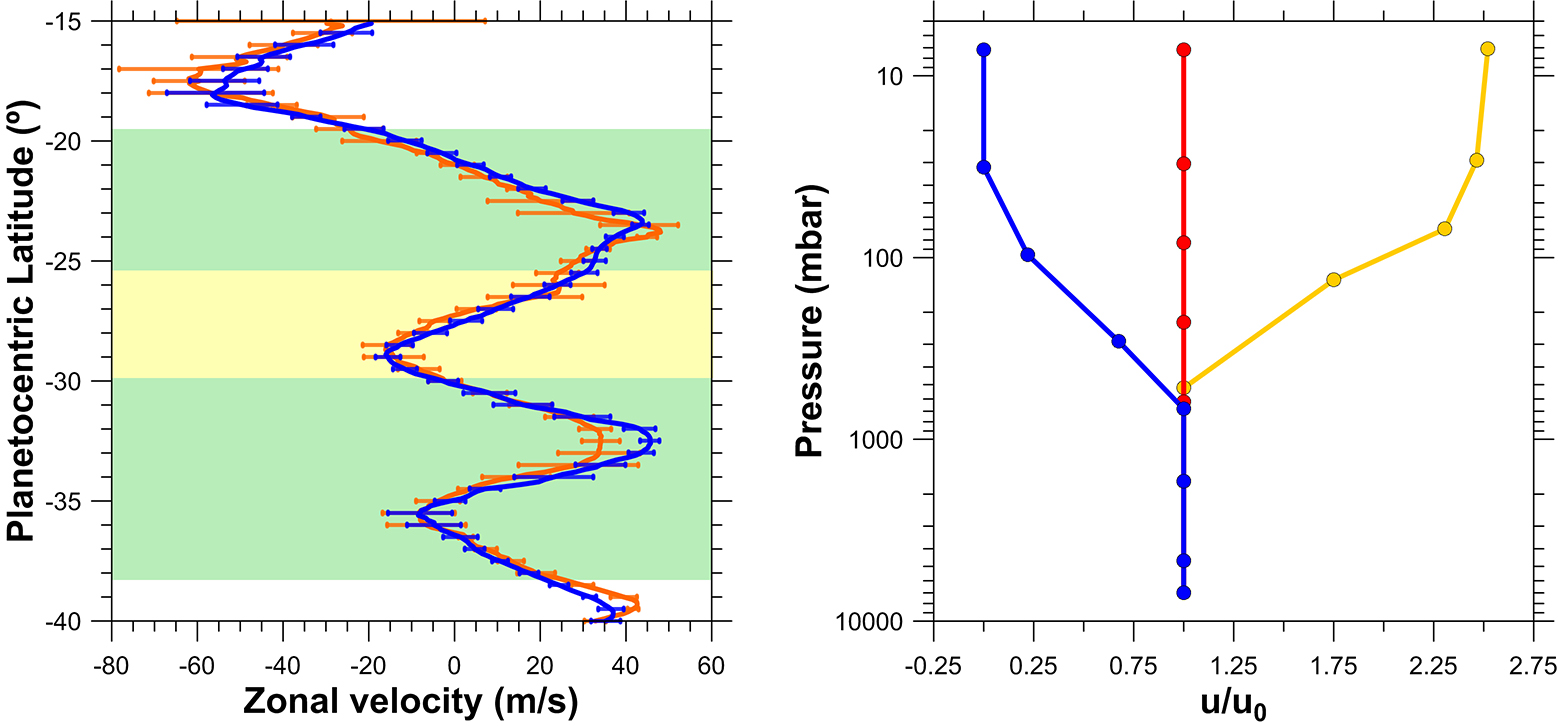}
	\caption{Zonal winds of the reference atmosphere. Left: zonal-wind profile at the cloud level. The orange profile on the zonal-wind profile is the profile measured with Cassini in 2000 \citep{Porco_Cassini_Jup_winds_2003}. The blue profile is the one measured in 2016 with HST images \citep{Hueso_Jupiter_before_Juno_2017}. The latitude range modelled in our simulations is highlighted with a light green box while the latitude range of the STB Ghost is highlighted with a light yellow box. Right: Amplitudes of the winds at different heights with respect to the zonal winds at cloud level ($u_0$). Blue line represents the nominal reference atmosphere, red represents an atmosphere with no vertical wind shear and yellow an atmosphere with winds increasing with altitude.}
	\label{figure EPIC atmosphere part2}
\end{figure}

The ``nominal" model atmosphere is equivalent to those used in \citet{Garcia-Melendo_2005, Garcia-Melendo_BA_I_2009} and \citet{Legarreta_EPIC_Jup_vortex_2008}, where a systematic exploration of the vertical wind shears above and below the clouds was performed resulting in a vertical structure of the winds similar to the one here used.

\subsection{Introducing vortices and storms}
Vortices are introduced in EPIC as a Gaussian ellipsoidal perturbation ($\Delta M$) of the Montgomery streamfunction $M=C_{p}T+gz$ \citep{Stratman_Neptune_EPIC_2001, Legarreta_EPIC_Jup_vortex_2008}. This variable plays the role of streamfunctions on fluids on geostrophic equilibrium following motions on isentropic surfaces. In EPIC the Montgomery streamfunction can be perturbed to create a region of unusual vorticity that will cause a vortex. The perturbation has the form

\begin{footnotesize}
\begin{equation}
    \Delta M = \alpha f R_{e} \, b_{s} \, V_{T} \, \exp \left\lbrace -\left[\left[ \left(\frac{\phi - \phi_{s}}{a_{s}}\right)^{2}+\left(\frac{\varphi - \varphi_{s}}{b_{s}}\right)^{2}\right]^{\frac{n}{2}} + \left( \frac{\ln(P)-\ln(P_{s})}{c_{s}} \right)^{2} \right]\right\rbrace
    \label{equation vortex}
\end{equation}
\end{footnotesize}

Here $f=2\Omega \sin\varphi_{s}$ is the Coriolis parameter, $\Omega$ is the planetary angular velocity, $R_{e}$ the equatorial radius and $V_{T}$ the tangential velocity of the vortex, with $V_{T}$ positive for anticyclones and negative for cyclones. $\phi$, $\varphi$ and $P$ are the east longitude, planetographic latitude and pressure, respectively, with the sub-index $s$ indicating the centre of the vortex and $P_{s}$ being the central pressure level where the vortex is injected. The size of the vortex is given by the semi-major axes $a_{s}$ and $b_{s}$ and its vertical extent is measured in scale-heights $c_{s}$. This vertical extent can be defined differently for the layers above and below the altitude of the vortex defining two quantities $c_{up}$ and $c_{down}$.

The distribution of velocities in the vortex is given by the non-dimensional parameter $\alpha$, which depends on a shape-factor $n$ as:
\begin{equation}
\alpha=\frac{\exp(1-\frac{1}{n})}{n(1-\frac{1}{n})^{1-1/n}}
\end{equation}

In these equations larger values of $n$ correspond to vortices with velocities more concentrated in an annular structure.

Therefore, a vortex is defined by 9 parameters: the centre of the vortex given by the three position coordinates ($\phi_{s}$, $\varphi_{s}$ and $P_{s}$), the longitudinal and latitudinal semi-major axes ($a_{s}$ and $b_{s}$), the vertical lengths above and below the pressure level where the vortex is injected ($c_{up}$ and $c_{down}$), the tangential velocity ($V_{T}$) and the shape-factor ($n$). As a result of the geostrophically balanced injection of the vortex, it is necessary to have a period of adjustment to balance the atmosphere, which for anticyclones usually lasts around one vortex turnaround time but can be much longer for cyclones.

Storms are introduced via heat pulses $\dot{Q}$ of Gaussian shape following \cite{Garcia-Melendo_2005}:

\begin{equation}
    \dot{Q} = \dot{Q}_{0} \exp \left(-\left(\frac{(\phi-\phi_{p})^{2}}{2 a^{2}_{p}} + \frac{(\varphi-\varphi_{p})^{2}}{2 b^{2}_{p}}\right)\right)
    \label{equation heat}
\end{equation}

where $\dot{Q}_{0}$ is the amplitude of the heating perturbation given in W$\,$kg$^{-1}$, $\phi$ is the east longitude, $\varphi$ is the planetographic latitude and $(\phi_{p}, \varphi_{p})$ are the initial coordinates of the heat pulse. The size of the pulse is given by the longitudinal and latitudinal semi-major axes $a_{p}$ and $b_{p}$ respectively. The pulses are introduced at each vertical layer of the model except at the abyssal layer.

In this work we have modified the EPIC code to have a better control of the pulses, indicating the start time and duration of each one. We have injected three pulses sequentially. Each pulse drifts longitudinally with a given velocity and is defined by 8 parameters: initial longitude and latitude, zonal drift rate, longitudinal and latitudinal extensions defining the shape of the pulse, the start time and duration of the pulse, which come from the observations of the three convective cores, and a constant heating amplitude over that time interval.

\subsection{Exploration of the parameter space}

Initiating elongated cyclones like the STB Ghost is not straightforward in EPIC. A single elongated cyclone tends to experience several changes in size as it expands and contracts before reaching a stable shape only after tens of days.  An alternative is to introduce chains of small circular cyclonic vortices that merge together to form an STB Ghost-like feature. However, the resulting vortex also experiences significant changes in size and shape before it stabilizes into an elongated cyclone. In both cases the final properties of the cyclone depend on parameters such as the tangential velocities, shape-factor $n$, and vertical structure ($c_{up}$, $c_{down}$).

\begin{table}[h]
\centering
\begin{tabular}{|l|c|c|}
\hline
    \textbf{Parameter}               & \textbf{Single  vortex}                      & \textbf{Multiple vortices}                    \\ \hline
    $\varphi_{pc}$ ($^{\circ}$)  & $-28.4$ $\longleftrightarrow$ $-25.9$   & $-27.5$ $\longleftrightarrow$ $-26.3$  \\ \hline
    $P_{s}$ (mbar)                    & $680$                                            & $680$                                              \\ \hline
    $a$ ($^{\circ}$)                   & $13$ $\longleftrightarrow$ $7$         & $2$ $\longleftrightarrow$ $4$             \\ \hline
    $b$ ($^{\circ}$)                  & $2$ $\longleftrightarrow$ $3$            & $1.25$ $\longleftrightarrow$ $2.9$      \\ \hline
    $c_{up}$ (scale-heights)      & $1$ $\longleftrightarrow$ $3$        & $1$ $\longleftrightarrow$ $3$              \\ \hline
    $c_{down}$ (scale-heights)  & $1$ $\longleftrightarrow$ $3$           & $1$ $\longleftrightarrow$ $3$              \\ \hline
    $V_{T}$ (m/s)                     & $30$ $\longleftrightarrow$ $120$    & $40$ $\longleftrightarrow$ $90$        \\ \hline
    $n$                                    & $2$, $3$                                        & $2$                                                   \\ \hline  
    Number of vortices                &            1                                      & $3$, $4$, $6$, $7$, $8$, $12$               \\ \hline
    $\Delta \lambda_{V}$ ($^{\circ}$) &        N.A.                            & $2$ $\longleftrightarrow$ $10$              \\ \hline
\end{tabular}
\caption{Explored parameter space to model the STB Ghost in EPIC. The parameters are: $\varphi_{pc}$: planetocentric latitude, $P_{s}$: pressure level where the vortexes are inserted, $a$: longitudinal semi-major axis, $b$: latitudinal semi-major axis, $c_{up}$: vortex upward altitude, $c_{down}$: vortex downward altitude, $V_{T}$: tangential velocity of the vortex, $n$ shape parameter and $\Delta \lambda_{V}$: separation between the centre of the vortices when initiating a system of multiple vortices.}
\label{tabla parameters Ghost}
\end{table}

Table \ref{tabla parameters Ghost} shows the range of parameters explored to simulate the Ghost, either as a unique vortex, or as a chain of vortices. A systematic exploration of the space of parameters was not done due to the large time scales required for the initial perturbation to evolve into a stable STB Ghost-like feature. Instead, the parameters have been tested by approximation, i.e., we tested reasonable values of the parameters based on the observations and we introduced small changes to the parameters observing whether the outcome generated a better match with the observations or not. The success of a simulation was determined by the comparison between the maps of simulated potential vorticity and the observed cloud field, and in particular looking at the final size and stability of the simulated Ghost, which were markedly different from the observations in unsuccessful simulations. The drift rate of the simulated Ghost was also taken into account, with the successful simulations matching the drift rate of the STB Ghost in the observations. However this parameter was almost fixed by choosing the right latitude for the Ghost and did not vary significantly by changing the depth or intensity of the Ghost. In our simulation, we changed only one of the parameters, while the others remained fixed. This way the effect of a single change could be observed to determine if it was a positive change or not in terms of finding a stable Ghost feature that resembled its behaviour in the observations. 

About 120 simulations were launched for the cyclone to test its latitude, circulation, shape-factor, size and single or multiple origin. The Ghost is left to evolve freely until the perturbations generated when the Ghost is inserted are dissipated and the vortex acquires a regular constant shape similar to the one observed. This stabilization typically requires 40-70 days, which is a time scale much larger than in previous studies of anticyclones in the jovian atmosphere, where the stabilization time roughly scales with the time it takes for a full revolution of the material in the vortex \citep{LeBeau_Neptune_GDS_1998, Morales_Cylones_Jupiter_2002, Legarreta_EPIC_Jup_vortex_2008}.  

Once we created a successful cyclone in terms of its size and morphology in potential vorticity maps and a stable shape and drift rate in agreement with the observations, we subsequently checked its behaviour when initiating it using a modified atmosphere (i.e. testing the HST 2016 constant-winds scenario, the Cassini 2000 constant winds scenario or scenarios with winds decaying or increasing in altitude above the cloud level). In general, the best simulations were obtained for the nominal atmosphere with the HST 2016 zonal winds assuming decaying winds above the visible cloud layer and constant winds below it. Attempts to run models of the Ghost with winds increasing in height above the clouds produced unstable simulations or vortices with drift rates very different from the observed drift rate of the Ghost.

Simulations favoured cyclones initiated as a single vortex. Chains of vortices merged quickly but produced stronger final cyclones that seem to have stronger circulations than in the observations. Single vortices with tangential velocities of $80\,$ms$^{-1}$ (similar to measurements on JunoCam images) were more successful in terms of the final shape and size of the elongated cyclone than weaker (like those based on HST images) or stronger circulations. Weaker circulations resulted in vortices with edges that were not well defined, and stronger circulations resulted in cyclonic vortices that were too elongated with respect to the observed Ghost.

Regarding the vertical extension of the cyclone, vertically narrow vortices were very unstable and resulted in errors in the simulations with isentropes crossing or in vortices that split and merged continuously. Vortices had to extend at least over $3$ scale-heights but the best simulations were obtained for vortices that extended vertically $3$ scale heights above the cloud top and $2$ scale heights below. With these numbers, and the exponential decay of the vortex properties assumed in Equation \ref{equation vortex}, our favoured Ghost cyclone decays by a factor of $e$ at the 5-bar level. We point out that this ``favoured" vertical extension makes the deeper part of the Ghost to just reach the water condensation level for a solar abundance of water.

Finally, a shape factor $n=2$ produced better results and produced a velocity distribution with low internal velocities and intense motions concentrated at the edges of the cyclone that favoured the consistency of the elongated cyclone.

\begin{table}[htbp]
\centering
\begin{tabular}{|l|c|}
\hline
    \textbf{Parameter}                & \textbf{Oval BA}                  \\ \hline
    $\varphi_{pc}$ ($^{\circ}$)   & $-29.6$ $\longleftrightarrow$ $-30.3$ \\ \hline
    $P_{s}$ (mbar)                     & $680$ \\ \hline
    $a$ ($^{\circ}$)                   & $3.0$ $\longleftrightarrow$ $4.85$ \\ \hline
    $b$ ($^{\circ}$)                   & $3.0$ $\longleftrightarrow$ $4.5$ \\ \hline
    $c_{up}$ (scale-heights)       & $3$ \\ \hline
    $c_{down}$ (scale-heights)   & $3$ \\ \hline
    $V_{T}$ (m/s)                      & $100$, $110$, $120$ \\ \hline
    $n$                                     & $2$, $3$ \\ \hline  
\end{tabular}
\caption{Explored parameter space to model the Oval BA in EPIC. The parameters are: $\varphi_{pc}$: planetocentric latitude, $P_{s}$: pressure level where the vortexes are inserted, $a$: longitudinal semi-major axis, $b$: latitudinal semi-major axis, $c_{up}$: vortex upward altitude, $c_{down}$: vortex downward altitude, $V_{T}$: tangential velocity of the vortex and $n$ shape parameter.}
\label{tabla parameters BA}
\end{table}

Once the STB Ghost acquired a fixed shape, Oval BA was inserted east of the Ghost as an anticyclone separated by $17^{\circ}-24^{\circ}$. The tangential velocities and vertical extension of Oval BA are based on wind measurements by \citet{Hueso_2009_BA} and numerical simulations with EPIC by \citet{Garcia-Melendo_BA_I_2009}. The range of the explored parameters used to simulate the anticyclone Oval BA can be seen in Table \ref{tabla parameters BA}. We launched 30 simulations of Oval BA covering these parameters. Most of these simulations were considered as successful when comparing simulations with Oval BA, since this anticyclone is highly constrained by the observations and previous works \citep{Garcia-Melendo_BA_I_2009}. 

\begin{table}[h]
\centering
\small{\begin{tabular}{|l|l|l|l|}
\hline
    \textbf{Parameter}              & \textbf{Storm 1}                              & \textbf{Storm 2}                              & \textbf{Storm 3}                      \\ \hline
    $\varphi_{pc}$ ($^{\circ}$) & $-27.4$ $\longleftrightarrow$ $-27.8$ & $-26.0$ $\longleftrightarrow$ $-26.9$ & $-27.2$ $\longleftrightarrow$ $-27.7$ \\ \hline
    $v$ (m/s)                           & $8.0$ $\longleftrightarrow$ $16.0$     & $-24.8$ $\longleftrightarrow$ $8.0$    & $10.0$ $\longleftrightarrow$ $14.0$   \\ \hline
    $a$ ($^{\circ}$)                 & $0.45$ $\longleftrightarrow$ $1.5$     & $0.35$ $\longleftrightarrow$ $1.25$    & $0.6$ $\longleftrightarrow$ $0.9$     \\ \hline
    $b$ ($^{\circ}$)                 & $0.35$ $\longleftrightarrow$ $0.6$     & $0.17$ $\longleftrightarrow$ $1.0$     & $0.35$ $\longleftrightarrow$ $0.6$    \\ \hline
    Time active (day)                & $4$ $\longleftrightarrow$ $6$            & $1$ $\longleftrightarrow$ $2$            & $2$ $\longleftrightarrow$ $3$         \\ \hline
    Start time (day)                  & $58$ $\longleftrightarrow$ $90$       & $65.5$ $\longleftrightarrow$ $90$    & $85.3$ $\longleftrightarrow$ $93$     \\ \hline
    $\dot{Q}_{0}$ (W/kg)         & $0.1$ $\longleftrightarrow$ $1.5$      & $0.3$ $\longleftrightarrow$ $0.8$       & $0.2$ $\longleftrightarrow$ $0.75$    \\ \hline
\end{tabular}}
\caption{Explored space of parameters of the heat pulses. The parameters are: $\varphi_{pc}$: the planetocentric latitude, $v$: the drift rate of the pulse, $a$: the longitudinal semi-major axis, $b$: the latitudinal semi-major axis, Time active: the number of days the pulse is kept active, Start time: the day when the pulse activates, and $\dot{Q}_{0}$ the amplitude of the heating perturbation.}
\label{tabla parameters Storms}
\end{table}

Depending on the particular simulation and stabilization times, at days 58-90 the heat pulses were introduced. We launched simulations that tested the intensity and size of the heat pulses, their latitudinal and longitudinal position and the duration of the convective activity for each pulse. We also launched simulations with only one, two or three heat pulses. A total of about 275 simulations were tested in this stage. The sizes of the heat pulses correspond to the sizes of the storms on the day each one started, except for the second storm, whose first observation was in relatively low quality images. In that case we used the size of the apparent convective core observed on 7 February 2018 in HST images. The heat pulses introduced in the simulations were activated following drift rates equal to those of the observed storms (Figure \ref{figure STB storms tracking}). Table \ref{tabla parameters Storms} summarizes the explored parameters to describe these storms. Most of the simulations resulted in maps of the potential vorticity that largely separated from the observations. In those unsuccessful cases, the interaction of the simulated storms with the cyclone did not generate a morphology similar to the observed one. Smaller or weak storms had a very small impact in the morphology of the Ghost and large or very intense storms completely tore apart the cyclone in a time-scale of a few days. Only the ones that reproduced the observations best were left to evolve for long periods of time. 

\subsection{Best simulation}

The selection of the best simulation was based on the comparison between the potential vorticity maps and the observed cloud field. Our best simulation was found using the nominal reference atmosphere. The STB Ghost was best reproduced with a single elongated unique vortex and Oval BA was best generated with values similar to those by \citet{Garcia-Melendo_BA_I_2009}. We also found that three convective storms better reproduced the morphology of the STB Disturbance than any combination we attempted of one or two convective events. This was particularly true when comparing the outcome of the simulations with the HST and JunoCam observations obtained three days after the onset of the convective activity. The parameters that define our best simulation are detailed in Tables \ref{tabla best parameters Ghost BA} (vortices) and \ref{tabla best parameters Storms} (storms).

\begin{table}[htbp]
\centering
\begin{tabular}{|l|l|l|l|}
\hline
    \textbf{Parameter}              & \textbf{STB Ghost}  & \textbf{Oval BA}   \\ \hline
    $\varphi_{pc}$ ($^{\circ}$) & $-27.3$                & $-30.2$            \\ \hline
    $P_{s}$ (mbar)                   & $680$                    & $680$              \\ \hline
    $a$ ($^{\circ}$)                  & $11.0$                   & $3.5$              \\ \hline
    $b$ ($^{\circ}$)                  & $2.3$                    & $3.5$              \\ \hline
    $c_{up}$ (scale-heights)      & $3$                       & $3$                \\ \hline
    $c_{down}$ (scale-heights)  & $2$                       & $3$                \\ \hline
    $V_{T}$ (m/s)                     & $-80$                    & $100$              \\ \hline
    $n$                                    & $2$                       & $2$                \\ \hline
\end{tabular}
\caption{Parameters for the modelled STB Ghost and Oval BA in the simulation that best fitted the observations.}
\label{tabla best parameters Ghost BA}
\end{table}

\begin{table}[htbp]
\centering
\begin{tabular}{|l|l|l|l|}
\hline
    \textbf{Parameter}              & \textbf{Storm 1}   & \textbf{Storm 2}  & \textbf{Storm 3} \\ \hline
    $\varphi_{pc}$ ($^{\circ}$) &  $-27.5$              & $-26.8$               & $-27.5$          \\ \hline
    $v$ (m/s)                           &  $14.0$               & $-12.0$               & $11.0$           \\ \hline
    $a$ ($^{\circ}$)                 &  $0.8$                 & $0.4$                  & $0.7$            \\ \hline
    $b$ ($^{\circ}$)                 &  $0.5$                 & $0.3$                  & $0.35$           \\ \hline
    Start time (day)                  &  $86.0$               & $88.0$                 & $88.5$           \\ \hline
    Time active (day)                &  $4.0$                 & $2.0$                  & $2.5$            \\ \hline
    $\dot{Q}_{0}$ (W/kg)         &  $0.6$                & $0.5$                   & $0.4$            \\ \hline
\end{tabular}
\caption{Parameters for the modelled convective storms in the simulation that best fitted the observations.}
\label{tabla best parameters Storms}
\end{table}

\begin{figure}[h]
	\centering
		\includegraphics[angle=0, width=1.0\textwidth]{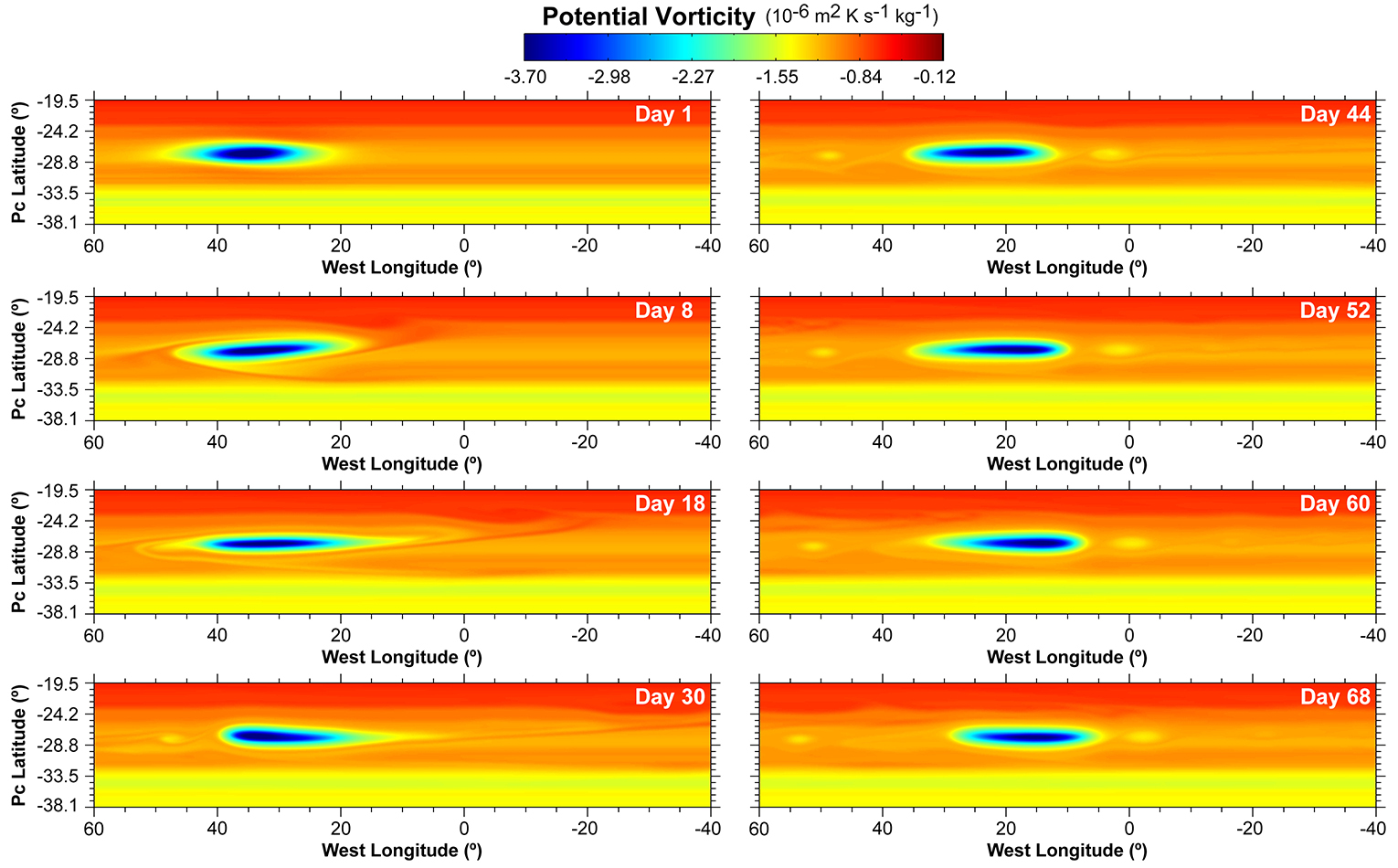}
	\caption{Potential vorticity maps at the $649\,$mbar pressure level showing the best simulation of the STB Ghost obtained with the EPIC model. The simulated Ghost was introduced as a single elongated vortex. Only a small portion of the model domain is represented in this figure.}
	\label{figure EPIC best Ghost}
\end{figure}

The Ghost was left to run freely during 68 days of stabilization. Figure \ref{figure EPIC best Ghost} shows the generation of the STB Ghost and how it expanded and contracted several times before acquiring a constant shape. At that time, Oval BA was introduced separated by $19.2^{\circ}$ from the east side of the simulated Ghost. The simulation was then left to evolve through another 18 days. At this point both structures were separated by $17.2\,^{\circ}$ and had drift rates and sizes similar to those observed on 2018 February 4. This was a result of the selection of appropriate latitude, size and circulation of the Ghost and Oval BA described in Table \ref{tabla best parameters Ghost BA}.

\begin{figure}[htbp]
	\centering
		\includegraphics[angle=0, width=0.60\textwidth]{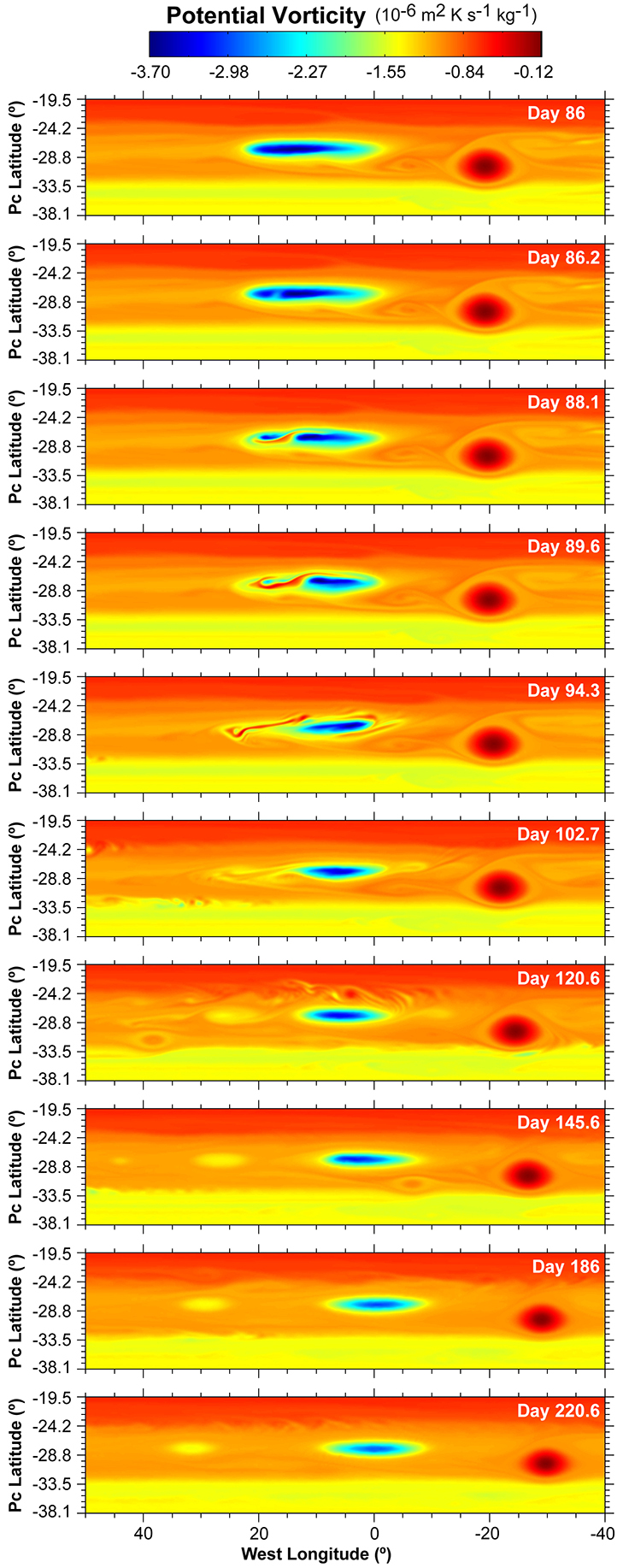}
	\caption{Potential vorticity maps at the $649\,$mbar pressure level showing the evolution of the best simulation of the STB Disturbance obtained with the outbreak of three convective storms. Only a small portion of the model domain is represented in this figure.}
	\label{figure EPIC best simulation}
\end{figure}

Then, we activated three heat sources on days $86$, $88$ and $88.5$. Each heat pulse had an intensity close to $0.5\,$Wkg$^{-1}$ and was active during 2-4 days. Figure \ref{figure EPIC best simulation} shows frames of the evolution of the potential vorticity field. The layer represented has potential temperature $\theta=187.5\,$K and average pressure $P=649\,$mbar, corresponding to the pressure level where the visible ammonia clouds are located. The first frame shows the simulated STB Ghost and Oval BA just before inserting the first heat pulse. The following frames show how the heat pulses produced a recirculation pattern that resembles the characteristic ``S"\ shape. About 8 days after the start of convection a recirculation pattern in the east side of the Ghost resembles the dark features circulating around the Ghost at that time. As a net result of the outflow produced by the convective perturbations, the inner circulation of the Ghost partially breaks and the material located south of the convective storms does not circulate inside the Ghost, but is expelled westward and drifted by the zonal winds.

Later stages of the evolution of the disturbance are not so well captured. Our simulations did not show the conversion of the STB Ghost from a quiescent state to a turbulent one. We believe that this could be due to insufficient spatial resolution to generate the very small-scale structures and the plausible presence of convective activity at later dates (as hinted by images on the methane absorption band from HST on 17 April 2018 and from PlanetCam UPV/EHU on 22 May 2018). The last frames show the accumulated effects of numerical dissipation limiting the total time that can be simulated to around 200 Earth days (480 jovian days). Since the effects of numerical dissipation depend on the spatial resolution of the simulations, longer simulations could be run at higher spatial resolutions but would require considerably more computation time.

As a summary of this section we can say that EPIC simulations successfully reproduce the effects of convective storms developed inside the elongated cyclone and how this interacts with Oval BA. The net effect of the storms in the simulations is to split the cyclone producing the expulsion of material from the Ghost that drifts to the west. There are some limitations in the model that are worth mentioning. The model is not able to generate spontaneously the vortices and storms and cannot explore whether the decreasing distance between Oval BA and the Ghost played a role to trigger the convective activity. Instead, all of the actors in the dynamics of this complex phenomenon are introduced in the model manually but the outcome of the simulation critically depends on their assumed characteristics. These characteristics have consequences that we now discuss.

\section{Discussion} \label{section_Discussion}

\cite{Dowling_1989_Cyclones_convection_Jupiter} proposed that the relation between cyclonic regions and moist convection in Jupiter could be explained by simple arguments of geostrophy. Cyclones are regions of low pressure that when expressed in surfaces of constant potential temperature produce a depression of the upper layers together with a rise in the deep layers. The simulated STB Ghost has a deep structure that extends down to just above the water condensation level at 5 bar. This invites us to speculate that a small perturbation of the cyclone may be caused by its close approach to Oval BA and the related atmospheric systems nearby. This proximity might serve to initiate moist convection at the bottom of the cyclone. The observational fact that the STB Ghost radically altered its drift rate at the time it approached Oval BA and the convective storms started suggests a relation between the three systems. However, the storms originated in the west side of the cyclone and not the east side where the interaction occurred. A model like EPIC is not well suited to study this question because it cannot introduce the release of latent heat directly, and convective storms need to be injected manually in the model.

\begin{figure}[htbp]
	\centering
		\includegraphics[angle=0, width=1.0\textwidth]{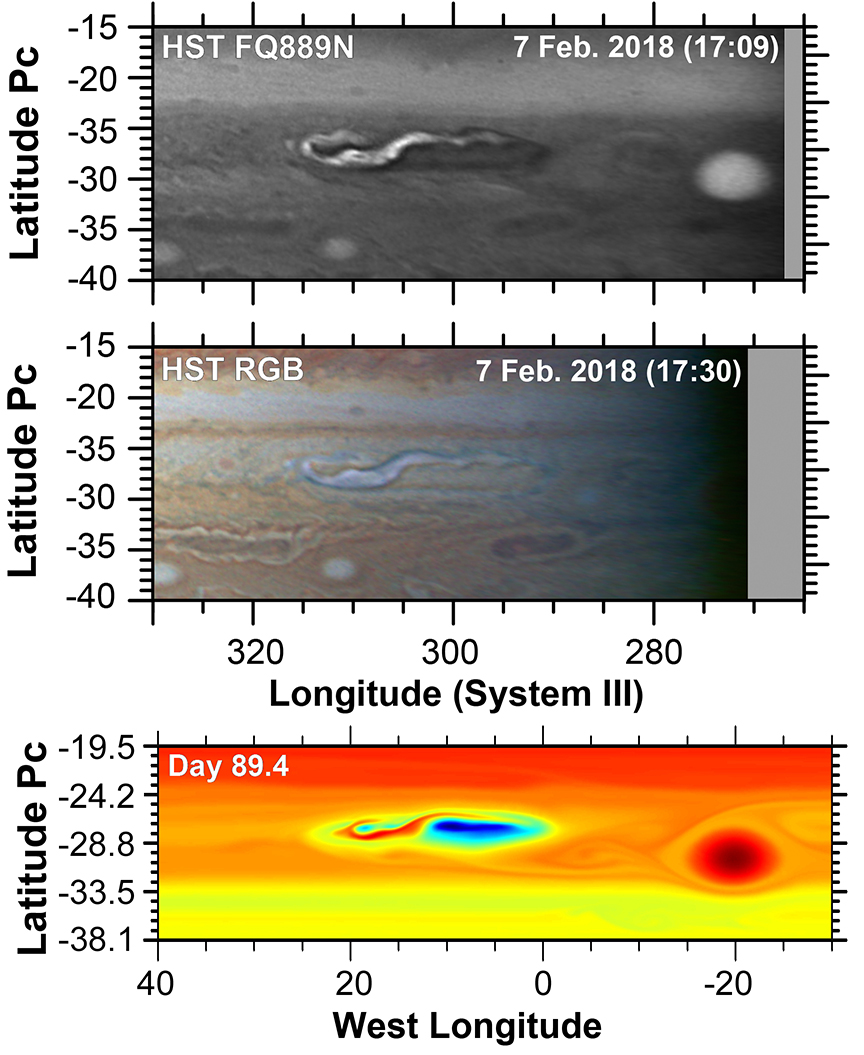}
	\caption{Comparison between the observations taken on 7 Feb. 2018 by HST and the potential vorticity map of our best simulation corresponding to the same stage of the disturbance. The HST colour composition map was generated using images captured with the F631N, F502N and F395N filters.}
	\label{figure EPIC comparison}
\end{figure}

Figure \ref{figure EPIC comparison} shows a comparison between the HST observations on 7 February 2018, around 3 and a half days after the onset of the first storm, and the equivalent potential vorticity map of the best simulation run with EPIC with storms that were manually injected in the simulation. The model reproduces well this first stage of the phenomenon, showing a similar morphology to the observed one with the first storm developing the characteristic ``S"\ shape and recirculating material at the west edge of the Ghost.

Moist convection powered by ammonia or water condensation should have very different behaviour due to the different amounts of energy that can be released when condensing these volatiles. We can estimate maximum temperature differences $\Delta T$ between an updraft and its environment:

\begin{equation}
    \Delta T=\chi_i \frac{\mu_i}{\bar{\mu}} \frac{L_i}{C_{p}}
\end{equation}

where $\chi_i$ is the molar abundance of water or ammonia, $\mu_i$ is the molecular weight of water ($18.02\,$gmol$^{-1}$) or ammonia ($17.03\,$gmol$^{-1}$), $\bar{\mu}=2.2\,$gmol$^{-1}$ is the mean molecular weight of the atmosphere, $L_i$ is the latent heat of condensation of water ({$L_{H_{2}O}=2,834\,$Jg$^{-1}$) or ammonia ($L_{NH_3}=1,836\,$Jg$^{-1}$) \citep{ICT} and $C_{p}=12,360\,$Jkg$^{-1}$K$^{-1}$ is the specific heat of the lower atmosphere considering an intermediate ortho-para hydrogen distribution. Considering a concentration of water equal to solar abundance, $\chi_{H_{2}O}=9.3\cdot 10^{-4}$ \citep{Asplund_2009_Chemical_composition_Sun}, and a deep concentration of ammonia $\chi_{NH_{3}}=2.4\cdot 10^{-4}$ from recent Juno results by  \cite{2017_GRL_Li_Ammonia_Juno} (roughly $1.9$ solar), then we should expect maximum temperature differences in the updraft with their environment of $\Delta T$ of $1.7\,$K and $0.3\,$K for water and ammonia moist convection, respectively.

Knowing the temperature difference between the heated parcel and the surroundings, the maximum vertical velocity of the updrafts can be estimated from the value of Convective Available Potential Energy (CAPE).

\begin{equation}
    CAPE=\int_{z_{1}}^{z_{2}} g_{eff} \frac{\Delta T}{T(z)} dz = \frac{w_{max}^{2}}{2},
\end{equation}
where $T(z)$ is the environment vertical temperature profile and $g_{eff}=23.6\,$ms$^{-2}$ is the effective acceleration of gravity at the STB. Then, the maximum vertical velocity in an updraft can be roughly estimated as

\begin{equation}
    w_{max} \approx \sqrt{2g_{eff}\frac{\Delta T}{<T>}h},
\end{equation}
where $h$ is the vertical length of the path followed by the ascending parcel and $<T>$ is the mean temperature.

The results of these calculations are $205\,$ms$^{-1}$ for water and $40\,$ms$^{-1}$ for ammonia. These values represent absolute maximums for the updrafts in the storms and are extreme because the calculation based on CAPE neglects the effects of mixing and entrainment, weight of the condensates and nonhydrostatic pressures. We have also not included effects of the static stability of the upper troposphere. However, the calculations above show that moist water convection with a deep abundance of solar water is potentially 25 times more energetic than moist ammonia convection.

We can compare these energy expectations to the energy injected into the model in our simulations. Integrating equation \ref{equation heat} and taking into account the mass per surface unit, the total amount of energy introduced in the atmosphere is:

\begin{equation}
    E = 2 \pi \, \dot{Q}_{0} \, a_{p} \, b_{p} \, \frac{P_{max}-P_{min}}{g_{\scriptscriptstyle{eff}}\,}\Delta{t},
\end{equation}
where $P_{max}$ and $P_{min}$ are the maximum and minimum pressures of the modelled atmosphere, $\Delta{t}$ is the total time of activity for the storm, and the $2\pi$ factor comes from the Gaussian shape of the heat pulse. Here $P_{max}=7.0\,$bar and $P_{min}=0.01\,$bar are the upper and lower bounds of the model.

For our successful simulations, table \ref{tabla energy storms} shows the range of intensities for each heat pulse, the total power and energy released for each storm and the mass of water or ammonia that should condense.

\begin{table}[htbp]
\centering
\scriptsize{\begin{tabular}{|l|l|l|l|l|l|}
\hline
$\bm{Storm}$    & $\dot{\bm{Q}}_{\bm{0}}$ \textbf{(Wkg$^{-1}$)} & $\bm{P}_{\bm{w}}$ \textbf{(W)} & $\bm{E}$ \textbf{(J)} & \textbf{M\textsubscript{H$_{\bm{2}}$O} (kg)} & \textbf{M\textsubscript{NH$_{\bm{3}}$} (kg)} \\ \hline
    \textbf{1} & $0.5-0.7$  & $5.0-7.0 \cdot 10^{16}$ & $1.7-2.4 \cdot 10^{22}$ & $6.1-8.5 \cdot 10^{15}$ & $9.4-13.1 \cdot 10^{15}$ \\ \hline
    \textbf{2} & $0.5-0.7$  & $1.5-2.1 \cdot 10^{16}$ & $2.6-3.6 \cdot 10^{21}$ & $9.2-12.9 \cdot 10^{14}$ & $1.4-2.0 \cdot 10^{15}$ \\ \hline
    \textbf{3} & $0.4-0.45$ & $2.4-2.7 \cdot 10^{16}$ & $5.3-5.9 \cdot 10^{21}$ & $1.9-2.1 \cdot 10^{15}$ & $2.9-3.2 \cdot 10^{15}$ \\ \hline
\end{tabular}}
\caption{Range of heating amplitudes required in the simulations, energies released and amounts of water and ammonia required to provide that energy. Here $\dot{Q_{0}}$ is the heating amplitude, $P_{w}$ is the power released, $E$ is the total energy released and M\textsubscript{H$_{2}$O} and M\textsubscript{NH$_{3}$} are the masses of water and ammonia respectively that need to condense to release that energy.}
\label{tabla energy storms}
\end{table}

The atmospheric mass that would be implied in vertical motions carrying water or ammonia would be:

\begin{equation}
    M_{atm}=\frac{\bar{\mu}}{\mu_{i}}\frac{M_{i}}{\chi_{i}},
\end{equation}
where $M_{i}$ is the mass of water or ammonia calculated in Table \ref{tabla energy storms} for each storm. The results of these calculations are shown in Table \ref{tabla storms origin}.

\begin{table}[h]
\centering
\small{\begin{tabular}{|c|c|c|c|}
\hline
    \multicolumn{2}{|c|}{} & \textbf{Only H$_{\bm{2}}$O condensation} & \textbf{Only NH$_{\bm{3}}$ condensation} \\ \hline
    \multirow{3}{*}{$\bm{M_{atm}}$ \textbf{(kg)}} & \textbf{Storm 1} & $8.0-11.2 \cdot 10^{17}$ & $5.1-7.1 \cdot 10^{18}$ \\ \cline{2-4}
     & \textbf{Storm 2} & $1.2-1.7 \cdot 10^{17}$ & $7.6-10.7 \cdot 10^{17}$ \\ \cline{2-4}
     & \textbf{Storm 3} & $2.4-2.8 \cdot 10^{17}$ & $1.5-1.7 \cdot 10^{18}$ \\ \hline
     \multicolumn{2}{|c|}{$\bm{\Delta T}$ \textbf{(K)}} & $1.7$ & $0.3$ \\ \hline
     \multicolumn{2}{|c|}{$\bm{w_{max}}$ \textbf{(m}\textbf{s$^{-1}$)}} & $205$ & $40$ \\ \hline
\end{tabular}}
\caption{Scale analysis of the required sizes for the possible sources of the storms. $M_{atm}$ is the atmospheric mass in the updrafts needed to uplift water or ammonia for each storm, $\Delta T$ is the temperature difference between the heated parcel and the environment and $w_{max}$ is the maximum vertical velocity of the updrafts.}
\label{tabla storms origin}
\end{table}

If we suppose that the storms are made of updrafts that collect their mass from a deep reservoir of volatiles vertically extending around the cloud condensation level, the horizontal surface of that reservoir will be given by:

\begin{equation}
    S=\frac{M_{atm} \, g_{eff}}{\Delta P},
\end{equation}
where $S$ is the surface of the reservoir and $\Delta P$ the pressure difference between the lower and upper pressure levels of the reservoir.

For reasonable values of $\Delta P$ for water condensation (assuming most of the water that is updrafted is concentrated between $5$ and $7\,$bar so that updrafts can extend a bit below the water condensation level for a solar water abundance at $5\,$bar), we obtain that condensing all the water available in the deep reservoir in an area of $1.4-1.8 \cdot 10^{8}\,$km$^2$ (roughly equivalent to the area of a circle of radius $6,600-7,700\,$km) could release enough energy to explain the observed storms. This size range is only a fraction of the STB Ghost size. If we assume ammonia condensation and $\Delta P$ from $0.7$ to $1.5\,$bar, we need to condense all ammonia available in an area of $2.2-2.9 \cdot 10^{9}\,$km$^2$ (roughly equivalent to the area of a circle of radius $26,300-30,500\,$km). This size range is much larger than the size of the STB Ghost and is clearly incompatible with the observed size of the storms. These order-of-magnitude calculations imply that NH$_{3}$ condensation alone is not powerful enough to produce the storms observed. Instead, water condensation from a region smaller than the STB Ghost could power up the storms observed and we can conclude that the storms developing the STB Disturbance were powered by water condensation with abundances close to or larger than solar abundance. This conclusion is also in agreement with expectations based on the morphology of the storm in methane-band images where high clouds are observed, which requires high velocities to reach the stable levels of the upper troposphere \citep{Hueso_2001_3D_model}. 

A comparison with the energies calculated for previous jovian storms can be performed. \citet{Gierasch_2000_Nature_Galileo_water_storms} estimated a power release of at least $5\times10^{15}\,$W for storms observed by Galileo with deep lightning on the night-side of the planet at depths with pressures of 3 bars or higher, implying water powered moist convection. The numbers we calculated from the EPIC simulations of the STBD convective cores are 3-14 times higher, which requires much stronger convection. The storms analysed by Galileo were regular storms that are often observed west of the Great Red Spot, but the storms in the STBD seem more exceptional with high and extensive cloud tops much larger than those of typical storms in Jupiter, except those growing to planetary-scale disturbances.

We can also compare energies in Table \ref{tabla energy storms} with detailed simulations of moist convective storms in Jupiter. Using a three-dimensional model of moist convection and considering moist water convection with 1 solar deep abundance \citet{Hueso_2001_3D_model} obtained that a single updraft of $60\,$km in radius developing updrafts with vertical velocities of $60\,$ms$^{-1}$ would liberate a power of about $2.2 \cdot 10^{15}\,$W. Much less vigorous storms were also simulated for moist ammonia convection storms. The total power introduced by each of the storms in the STBD in our simulations with EPIC are 15-30 times bigger than those single-cell calculations. The total area covered by the storm in Figure \ref{figure STB storms tracking} is about 12 times as large as the individual convective storms identified.  \citet{Hueso_JGR_2002} proposed that the large-scale storms seen on Jupiter could be formed by multiple smaller single cells and proposed numbers of tens of these single cell updrafts to reproduce large moist convective events to explain disturbances in the South Equatorial Belt. We conclude that water moist convection with solar water abundance or higher, instead of ammonia is the most plausible energy source driving the storms in the STB Ghost.

\section{Summary and conclusions} \label{section_Conclusions}

The STB Ghost was an elongated coherent cyclone located at the South Temperate Belt that experienced the development of a major convective perturbation when it approached close to Oval BA in February 2018. Observations of the STB Ghost made with JunoCam before the start of the convective disturbance show high-speed flows with strong zonal shears of the meridional flow at the east and west edges, internal compact small vortices, groups of small-scale bright clouds and an undulating collar with low albedo surrounded by another peripheral collar with higher albedo. The 2018 Jupiter's South Temperate Belt Disturbance that started in the west side of this cyclone was triggered by the eruption of three convective storms between 4 and 7 February 2018. The bright clouds of the storms expanded and scattered following the Ghost internal wind field. The turbulence generated by the storms was initially contained inside the STB Ghost, and ended up perturbing the region for several months. As a result of the evolution of the phenomenon, a large dark tail was generated westward of the STB Ghost and was expelled to the west while the STB Ghost approached to the anticyclonic Oval BA. The Ghost interacted with a small cyclonic cell and a white oval to the west of Oval BA merging with the first one and being deformed by the second one, arriving finally to the west of Oval BA. Oval BA acted as a ``barrier" to the eastward zonal drift of the STB Ghost which decreased significantly. However, Oval BA did not modify its drift rate significantly. By the end of October, the activity seemed to have ended with a morphology similar to the one before the disturbance started, with Oval BA followed by a smaller cyclonic cell on its northwest side formed by the merger of the Ghost and the cyclonic cell west of Oval BA. Between October and December 2018 this cyclonic system elongated and developed a shape reminiscent of the STB Ghost accompanied by a reactivation of the turbulence without clear hints of new convective storms. At the time of this writing (April 2019) this structure is characterized by dark and bright filaments very similar to the disturbed Ghost one year before.

A detailed comparison between the cloud morphology of the existing observations with simulations with the EPIC model of the potential vorticity field results in the following conclusions:

\begin{itemize}

\item Simulations with the EPIC model reproduce the overall phenomenology assuming that the STB Ghost is a cyclone characterized by strong winds ($\sim 80\,$ms$^{-1}$) concentrated in its outer region and vertically extended 4-5 scale-heights (from $90\,$mbar to $5\,$bar, although the upper limit is not well constrained due to the use of a sponge-layer in the upper layers in EPIC). These intense winds are roughly compatible with measurements based on analyses of JunoCam images. Lower spatial resolution data sets like HST may not resolve the small-scale features in the outer ring moving at the fastest velocities.

\item The storms produced abundant turbulence in the STB Ghost, inducing the creation of another structure to the west that evolved over months. Several anticyclones were formed and expelled from the south branch of the STB Ghost as a consequence of this activity. This aspect is also reproduced by the numerical simulations.

\item In our simulations we had to introduce storms with energies on the order of $2.5 \cdot 10^{22}\,$J to reproduce most of the observed characteristics of the early stages of the disturbance. This amount of energy is compatible with storms powered by water moist convection and requires at least a solar water abundance. We note that the depth of the simulated Ghost that best reproduced the observations extended to the $5$-bar level just above the water condensation level in Jupiter and we speculate that a perturbation to the Ghost when it approached Oval BA and its drift rate diminished could have served as a trigger to the development of the convective storms. This argument, however, cannot be proved or explored with the EPIC model, where the onset of convective storms cannot be predicted.

\item Although the storms that developed the STBD were powerful, they did not trigger a planetary-scale disturbance. The storms that grow to develop large planetary-scale disturbances in the NTB and SEB must also be powered by water condensation. The comparison of expansion rates of the storms with other moist convective storms \citep{Hunt_Nature_1982_storms_growth, Hueso_Icarus_1998_Hot_spots} suggests that the energy source for most of the storms observed in the visible and methane absorption band images in Jupiter is latent heat release in water condensation.

\end{itemize}

\section*{Acknowledgements}
We are very grateful to two anonymous reviewers that provided constructive and insightful comments. We are very thankful to observers operating small telescopes that submit their Jupiter observations to databases such as PVOL and ALPO-Japan. Observations were also obtained from the Centro Astron\'omico Hispano Alem\'an (CAHA) at Calar Alto, Spain which at the time of the observations reported was operated jointly by the Max Planck Institut f\"ur Astronomie and the Instituto de Astrof\'isica de Andaluc\'ia (CSIC) and is now operated by the Junta de Andaluc\'ia and the Instituto de Astrof\'isica de Andaluc\'ia (CSIC). JunoCam images were possible by the development and operations of the JunoCam instrument by Malin Space Science Systems (MSSS). This work also used data acquired from the NASA/ESA HST Space Telescope and archived by the Space Telescope Science Institute, which is operated by the Association of Universities for Research in Astronomy, Inc., under NASA contract NAS 5-26555. These observations are associated with programs GO  14661, 14839, 14756, 14936 and 15262 and we are thankful to the P. I.s of these observations M. Wong, I. de Pater and A. A. Simon for obtaining these data. Maps from GO 14756 and 15262 part of the OPAL program are available at \url{http://dx.doi.org/10.17909/T9G593} and maps from GO 14661, part of the Wide Field Coverage for Juno program, are available at \url{https://archive.stsci.edu/prepds/wfcj/}. This work was supported by the Spanish MINECO project AYA2015-65041-P (MINECO/FEDER, UE), Grupos Gobierno Vasco IT-765-13, IT-1366-19 and ``Infraestructura" grants from Gobierno Vasco and UPV/EHU. P. I\~nurrigarro  acknowledges a PhD scholarship from Gobierno Vasco. C. J. Hansen and G. S. Orton were supported by funds from NASA to the Juno mission via the Planetary Science Institute and the Jet Propulsion Laboratory, California Institute of Technology, respectively.


\appendix
\twocolumn
\section{Amateur data details}
\setcounter{table}{0}
\begin{tiny}
\tablehead{
\hline
\textbf{Date}   & \textbf{Time} & \textbf{Observer}    & \textbf{Filter} \\
yyyy/mm/dd   & UT & &\\
\hline
}
\tabletail{\hline}
\bottomcaption{List of amateur observations used in this study.}
\begin{xtabular}{|p{1.3cm}|p{0.5cm}|p{1.5cm}|p{0.73cm}|}
\label{table_amateur}
2017/02/24 & 21:03 & T. Olivetti             & Colour \\ 
2017/02/25 & 06:38 & D. Peach                 & Colour \\ 
2017/03/07 & 07:30 & D. Peach                 & Colour \\ 
2017/03/12 & 05:25 & A. Garbelini            & Colour \\ 
2017/03/24 & 15:40 & C. Go               & Colour \\ 
2017/03/31 & 16:57 & T. Olivetti             & Colour \\ 
2017/04/08 & 23:01 & R. Bosman               & Colour \\ 
2017/04/10 & 14:30 & C. Go               & Colour \\ 
2017/04/15 & 21:38 & M. Kardasis               & Colour \\ 
2017/04/17 & 13:51 & C. Go               & Colour \\ 
2017/04/19 & 14:07 & C. Go               & Colour \\
2017/04/21 & 16:32 & T. Olivetti             & Colour \\
2017/04/24 & 13:29 & C. Go               & Colour \\ 
2017/04/26 & 13:52 & C. Go               & Colour \\ 
2017/04/29 & 12:15 & A. Wesley               & Colour \\
2017/05/01 & 13:05 & C. Go               & Colour \\ 
2017/05/18 & 12:04 & C. Go               & Colour \\ 
2017/06/11 & 12:07 & C. Go               & Colour \\
2017/06/11 & 21:57 & Pic du Midi Peach            & Colour \\ 
2017/06/13 & 13:38 & T. Olivetti             & Colour \\ 
2017/06/14 & 10:53 & C. Go               & Colour \\
2017/06/14 & 20:34 & M. Lewis                 & 642nm \\
2018/01/25 & 21:03 & C. Go               & Colour \\ 
2018/02/04 & 09:29 & D. Peach                 & Colour \\ 
2018/02/04 & 18:16 & A. Wesley               & IR 750 \\
2018/02/04 & 18:19 & A. Wesley               & Methane \\ 
2018/02/06 & 00:49 & C. Foster                 & IR 685 \\ 
2018/02/06 & 09:25 & D. Peach                 & IR \\ 
2018/02/06 & 19:25 & A. Wesley               & Methane \\
2018/02/06 & 19:38 & A. Wesley               & IR \\ 
2018/02/07 & 16:26 & A. Wesley               & IR 750 \\
2018/02/08 & 11:43 & Unknown                      & Colour \\ 
2018/02/08 & 21:23 & C. Go               & IR \\
2018/02/09 & 09:11 & D. Peach                 & Colour \\ 
2018/02/09 & 18:47 & A. Wesley               & IR \\ 
2018/02/10 & 23:25 & T. Olivetti             & IR \\ 
2018/02/11 & 18:36 & A. Wesley               & IR \\
2018/02/13 & 09:54 & D. Peach                 & Colour \\ 
2018/02/13 & 19:43 & P. Miles                   & IR 742 \\ 
2018/02/15 & 22:10 & T. Olivetti             & IR \\ 
2018/02/16 & 08:53 & D. Peach                 & Colour \\
2018/02/16 & 18:22 & J. Kazanas                 & Colour \\
2018/02/17 & 22:41 & T. Olivetti             & IR \\ 
2018/02/18 & 00:41 & C. Foster                 & Colour \\ 
2018/02/18 & 09:38 & D. Peach                 & Colour \\
2018/02/18 & 19:51 & A. Wesley               & Colour \\
2018/02/20 & 21:40 & C. Go               & Colour \\ 
2018/02/23 & 08:24 & A. Soares                 & Colour \\
2018/02/25 & 19:53 & D. P. Milika       & R \\ 
2018/02/27 & 21:16 & C. Go               & Colour \\ 
2018/02/28 & 08:00 & A. Soares                 & Colour \\ 
2018/02/28 & 20:02 & C. Go               & Colour \\
2018/03/01 & 03:06 & C. Foster                 & IR \\ 
2018/03/02 & 19:41 & C. Go               & Colour \\ 
2018/03/03 & 07:24 & A. Soares                 & Colour \\
2018/03/04 & 01:13 & C. Foster                 & IR \\ 
2018/03/04 & 20:37 & C. Go               & Colour \\ 
2018/03/05 & 06:26 & A. Soares                 & Colour \\
2018/03/05 & 08:28 & F. Carvalho               & Colour \\
2018/03/07 & 08:41 & D. Peach                 & IR 685 \\
2018/03/07 & 19:18 & C. Go               & Colour \\ 
2018/03/09 & 20:12 & C. Go               & Colour \\ 
2018/03/12 & 07:59 & F. Carvalho               & Colour \\
2018/03/14 & 18:58 & T. Horiuchi             & Colour \\ 
2018/03/16 & 02:29 & C. Foster                 & Colour \\
2018/03/17 & 15:42 & T. Tranter                 & Colour \\ 
2018/03/17 & 17:41 & T. Horiuchi             & Colour \\ 
2018/03/19 & 18:43 & A. Wesley               & Colour \\
2018/03/21 & 19:49 & C. Go               & Colour \\
2018/03/22 & 06:22 & A. Soares                 & Colour \\
2018/03/22 & 17:05 & L. Westerland             & Colour \\ 
2018/03/24 & 18:30 & A. Wesley               & Colour \\ 
2018/03/25 & 05:36 & W. Martins               & Colour \\ 
2018/03/26 & 19:19 & T. Kumamori             & Colour \\ 
2018/03/28 & 19:46 & K. Suzuki              & Colour \\ 
2018/03/29 & 07:51 & M. B. Sparrenberger & Colour \\ 
2018/03/29 & 16:42 & P. Miles                   & IR 700 \\
2018/03/31 & 18:08 & T. Horiuchi             & Colour \\ 
2018/04/01 & 14:53 & T. Tranter                 & Colour \\
2018/04/03 & 05:00 & W. Martins               & Colour \\
2018/04/03 & 16:37 & C. Go               & Colour \\
2018/04/05 & 17:14 & A. Wesley               & Colour \\
2018/04/06 & 04:59 & A. Soares                 & Colour \\ 
2018/04/07 & 17:51 & A. Wesley               & IR750 \\ 
2018/04/08 & 14:17 & A. Casely                  & Colour \\ 
2018/04/10 & 15:29 & R. Iwamasa              & IR 685 \\ 
2018/04/12 & 18:08 & T. Olivetti             & Colour \\ 
2018/04/13 & 23:36 & C. Foster                 & Colour \\ 
2018/04/15 & 05:29 & E. Morales               & IR 685 \\ 
2018/04/19 & 00:54 & C. Foster                 & Colour \\ 
2018/04/20 & 04:42 & A. Soares                 & Colour \\ 
2018/04/20 & 14:32 & A. Casely                  & IR 642 \\ 
2018/04/22 & 16:07 & T. Horiuchi             & IR 685 \\
2018/04/23 & 02:27 & W. Martins               & Colour \\ 
2018/04/23 & 23:37 & C. Foster                 & IR 685 \\ 
2018/04/24 & 17:20 & C. Go               & Colour \\ 
2018/04/25 & 03:50 & W. Martins               & Colour \\ 
2018/04/27 & 16:03 & C. Go               & Colour \\
2018/04/28 & 21:01 & C. Foster                 & Colour \\
2018/04/29 & 16:22 & C. Go               & IR \\ 
2018/04/30 & 04:33 & L. A. Gomez            & Colour \\
2018/04/30 & 12:25 & A. Casely                  & IR 642 \\
2018/04/30 & 14:01 & T. Kumamori             & Colour \\
2018/05/01 & 17:25 & S. K. Chuen             & Colour \\ 
2018/05/01 & 20:51 & C. Foster                 & Colour \\ 
2018/05/02 & 14:43 & C. Go               & Colour \\
2018/05/04 & 05:08 & A. Coffelt                 & Colour \\ 
2018/05/04 & 16:14 & O. Inoue                  & Colour \\
2018/05/05 & 02:54 & D. Peach                 & Colour \\
2018/05/05 & 12:51 & L. Westerland             & Colour \\
2018/05/05 & 22:56 & M. Lewis                 & Colour \\
2018/05/07 & 03:28 & D. Peach                 & Colour \\
2018/05/07 & 04:12 & A. Soares                 & Colour \\
2018/05/07 & 04:59 & D. Peach                 & Colour \\
2018/05/07 & 13:59 & T. Tranter                 & Colour \\
2018/05/09 & 04:24 & D. Peach                 & Colour \\
2018/05/09 & 05:06 & E. Chappel                & Colour \\
2018/05/09 & 05:54 & M. Hood                    & Colour \\ 
2018/05/10 & 01:39 & L. A. Gomez            & Colour \\
2018/05/10 & 01:54 & D. Peach                 & Colour \\
2018/05/11 & 07:23 & G. Lamy                  & Colour \\ 
2018/05/11 & 16:09 & A. Yamazaki             & Colour \\ 
2018/05/12 & 04:24 & A. Coffelt                 & Colour \\ 
2018/05/12 & 12:18 & T. Kumamori             & Colour \\
2018/05/12 & 14:57 & R. Iwamasa              & Colour \\ 
2018/05/14 & 04:09 & E. Morales               & IR 685 \\ 
2018/05/14 & 05:17 & A. Coffelt                 & Colour \\
2018/05/14 & 14:22 & S. Ota                  & Colour \\ 
2018/05/15 & 00:46 & L. A. Gomez            & Colour \\
2018/05/17 & 12:54 & A. Yamazaki             & Colour \\
2018/05/19 & 13:13 & C. Go               & Colour \\ 
2018/05/19 & 23:56 & L. Martin                 & IR 642 \\ 
2018/05/21 & 03:29 & E. Morales               & IR 685 \\ 
2018/05/21 & 06:48 & G. Lamy                  & Colour \\ 
2018/05/21 & 15:31 & T. Yoshida             & Colour \\ 
2018/05/22 & 12:29 & A. Wesley               & IR 750 \\ 
2018/05/22 & 13:08 & C. Go               & Colour \\ 
2018/05/24 & 02:11 & D. Peach                 & Colour \\
2018/05/24 & 03:22 & D. Peach                 & Colour \\ 
2018/05/24 & 12:52 & C. Go               & Colour \\ 
2018/05/26 & 03:40 & D. Peach                 & IR \\ 
2018/05/26 & 06:12 & G. Lamy                  & Colour \\ 
2018/05/26 & 13:29 & C. Go               & Colour \\ 
2018/05/27 & 20:31 & A. Elia                 & Colour \\
2018/05/28 & 14:31 & M. Wong                 & IR 685 \\ 
2018/05/29 & 01:56 & P. Enache              & Colour \\ 
2018/05/29 & 01:56 & E. Morales               & Colour \\ 
2018/05/29 & 12:04 & A. Wesley               & IR \\ 
2018/05/31 & 03:21 & B. Macdonald              & Colour \\
2018/05/31 & 12:43 & C. Go               & Colour \\
2018/05/31 & 13:48 & C. Go               & Colour \\ 
2018/06/02 & 14:52 & T. Horiuchi             & Colour \\ 
2018/06/02 & 23:14 & M. Lewis                 & IR 642 \\ 
2018/06/03 & 01:54 & B. Macdonald              & Colour \\ 
2018/06/03 & 10:37 & A. Wesley               & IR 750 \\ 
2018/06/03 & 21:04 & M. Lewis                 & IR 642 \\ 
2018/06/04 & 05:30 & B. Macdonald              & Colour \\ 
2018/06/04 & 15:02 & S. Ota                  & Colour \\ 
2018/06/05 & 11:55 & A. Wesley               & IR 750 \\ 
2018/06/05 & 13:00 & S. Ota                  & Colour \\ 
2018/06/05 & 21:54 & M. Suarez                & Colour \\ 
2018/06/07 & 03:26 & G. Lamy                  & Colour \\ 
2018/06/07 & 12:45 & T. Yoshida             & Colour \\
2018/06/10 & 12:39 & L. Westerland             & Colour \\
2018/06/11 & 17:46 & C. Foster                 & IR685 \\
2018/06/12 & 02:31 & B. Macdonald              & Colour \\ 
2018/06/12 & 11:38 & C. Go               & Colour \\ 
2018/06/12 & 13:26 & A. Wesley               & IR 750 \\
2018/06/15 & 20:18 & A. Vilchez              & Colour \\
2018/06/19 & 03:19 & B. Macdonald              & Colour \\
2018/06/22 & 09:56 & A. Casely                  & Colour \\
2018/06/22 & 11:50 & A. Casely                  & IR \\
2018/06/23 & 08:46 & A. Wesley               & Colour \\
2018/06/24 & 11:54 & C. Go               & Colour \\ 
2018/06/26 & 14:21 & C. Go               & IR \\ 
2018/06/27 & 20:42 & M. Lewis                 & IR \\
2018/06/29 & 11:41 & C. Go               & Colour \\
2018/07/01 & 11:51 & S. Buda                  & Colour \\ 
2018/07/07 & 19:48 & J. L. Dauvergne           & Colour \\
2018/07/08 & 23:57 & P. Enache              & Colour \\
2018/07/09 & 09:34 & A. Casely                  & Colour \\ 
2018/07/09 & 11:38 & C. Go               & IR \\
2018/07/11 & 02:17 & B. Macdonald              & Colour \\
2018/07/11 & 10:50 & A. Wesley               & IR 685 \\
2018/07/14 & 07:30 & A. Wesley               & IR 750 \\
2018/07/14 & 09:33 & M. Wong                 & IR 685 \\ 
2018/07/16 & 12:36 & T. Kumamori             & Colour \\ 
2018/07/17 & 07:53 & A. Casely                  & IR 642 \\
2018/07/18 & 12:09 & T. Kumamori             & Colour \\ 
2018/07/21 & 09:12 & A. Casely                  & IR 642 \\
2018/07/21 & 10:41 & A. Casely                  & IR 642 \\
2018/07/23 & 11:31 & T. Kumamori             & Colour \\ 
2018/07/23 & 22:13 & P. Enache              & Colour \\ 
2018/07/28 & 12:30 & T. Kumamori             & Colour \\ 
2018/07/31 & 08:05 & A. Wesley               & IR 750 \\ 
2018/08/07 & 00:58 & B. Macdonald              & Colour \\
2018/08/09 & 00:26 & D. Peach                 & Colour \\
2018/08/12 & 16:38 & C. Foster                 & Colour \\ 
2018/08/13 & 22:21 & A. Soares                 & Colour \\
2018/08/14 & 08:26 & T. Barry                 & IR 685 \\
2018/08/15 & 16:00 & C. Foster                 & IR 685 \\ 
2018/08/16 & 22:47 & D. Peach                 & IR \\
2018/08/19 & 08:28 & T. Barry                 & IR 685 \\ 
2018/08/24 & 08:52 & A. Casely                  & Colour \\
2018/09/03 & 16:54 & C. Foster                 & IR \\ 
2018/09/05 & 08:01 & A. Casely                  & IR 642 \\
2018/09/10 & 16:29 & C. Foster                 & IR 685 \\ 
2018/09/13 & 15:54 & C. Foster                 & IR 685 \\
2019/01/26 & 19:03 & P. Miles                   & Colour \\
2019/02/02 & 10:07 & D. Peach                 & Colour \\
2019/02/18 & 02:47 & C. Foster                 & Colour \\
2019/02/25 & 02:41 & C. Foster                 & Colour \\ \hline
\end{xtabular}
\end{tiny}

\onecolumn

\onecolumn
\newpage

\setcounter{table}{0}
\section{Features drift rate}
\begin{footnotesize}
\tablehead{
\hline
\textbf{Name} & $\bm{\varphi_{pc}\,(^{\circ})}$ & \textbf{Drift rate}$\,$($\bm{^\circ}$\textbf{day}$^{\bm{-1}}$) & \textbf{u}$\,$(\textbf{ms}$\bm{^{-1}}$) & $\bm{\bar{u}}\,$\textbf{(ms}$\bm{^{-1}}$\textbf{)} \\ \hline
}
\tabletail{\hline}
\bottomcaption{Planetocentric latitude ($\varphi_{pc}$), drift rate and drift speed (u) of the features tracked during the disturbance compared with zonal winds ($\bar{u}$) at each from \citet{Hueso_Jupiter_before_Juno_2017}.}
\begin{xtabular}[h]{|c|c|c|c|c|}
\label{table drift rate features}
     Oval BA & $ -29.7 \pm 0.3 $ & $ -0.152 \pm 0.001 $ & $ 1.87 \pm 0.01 $ & $ 9 \pm 3 $ \\ \hline
     STB Ghost & $ -27.9 \pm 0.5 $ & $ -0.170 \pm 0.008 $ & $ 2.1 \pm 0.1 $ & $ -4 \pm 4 $ \\ \hline
     White oval & $ -30.3 \pm 0.3 $ & $ -0.112 \pm 0.002 $ & $ 1.37 \pm 0.02 $ & $ 3 \pm 4 $ \\ \hline
     Cyclonic cell & $ -27.5 \pm 0.4 $ & $ -0.12 \pm 0.01 $ & $ 1.5 \pm 0.1 $ & $ 3 \pm 4 $ \\ \hline
     Storm 1 & $ -27.7 \pm 0.2 $ & $ -1.1 \pm 0.1 $ & $ 14 \pm 1 $ & $ 0 \pm 4 $ \\ \hline
     Storm 2 & $ -26.4 \pm 0.3 $ & $ 1.0 \pm 2.5 $ & $ -12 \pm 10 $ & $ 18 \pm 4 $ \\ \hline
     Storm 3 & $ -27.6 \pm 0.2 $ & $ -0.9 \pm 0.3 $ & $ 11 \pm 4 $ & $ 1 \pm 4 $ \\ \hline
     Merged oval & $ -29.7 \pm 0.3 $ & $ -0.071 \pm 0.003 $ & $ 0.87 \pm 0.04 $ & $ -9 \pm 3 $ \\ \hline
     Dark tail (east) & $ -29.6 \pm 0.2   $ & $ -0.17 \pm 0.01 $ & $ 2.1 \pm 0.1 $ & $ -10 \pm 3 $ \\ \hline
     Dark tail (west) & $ -29.6 \pm 0.2   $ & $ 0.09 \pm 0.009 $ & $ -1.1 \pm 0.1 $ & $ -10 \pm 3 $ \\ \hline
     Dark spot 1 & $ -25.7 \pm 0.3 $ & $ -25.7 \pm 0.3  $ & $ 4 \pm 10     $ & $ 28 \pm 2 $ \\ \hline
     Dark spot 2 & $ -26.6 \pm 0.2 $ & $ -0.1 \pm 0.5   $ & $ 1 \pm 6      $ & $ 16 \pm 5 $ \\ \hline
     Dark spot 3 & $ -26.2 \pm 0.2 $ & $ -0.7 \pm 0.6   $ & $ 9 \pm 8      $ & $ 22 \pm 4 $ \\ \hline
     Dark spot 4 & $ -25.6 \pm 0.2 $ & $ -5 \pm 3 $ &   $ 64 \pm 39        $ & $ 29 \pm 3 $ \\ \hline
     Dark spot merged & $ -27.3 \pm 0.4 $ & $ 1 \pm 3 $ &   $ -13 \pm 38        $ & $ 6 \pm 4  $ \\ \hline
     Feature 1 & $ -28 \pm 0.2   $ & $ -0.0 \pm 0.2   $ & $ 0 \pm 3      $ & $ -6 \pm 4 $ \\ \hline
     Feature 2 & $ -28.2 \pm 0.4 $ & $ 0.2 \pm 0.2    $ & $ -3 \pm 3     $ & $ -8 \pm 4 $ \\ \hline
	 Feature 3 & $ -28.3 \pm 0.3 $ & $ -0.3 \pm 0.3   $ & $  4 \pm 4     $ & $ -10 \pm 4 $ \\ \hline
	 Feature 4 & $ -27.9 \pm 0.1 $ & $ -0.4 \pm 0.6   $ & $  5 \pm 8     $ & $ -4 \pm 4 $ \\ \hline
	 Feature 5 & $ -29.1 \pm 0.1 $ & $ -1.0 \pm 1     $ & $ 12 \pm 12    $ & $ -15 \pm 3 $ \\ \hline
 	 Middle fragment & $ -27.4 \pm 0.4 $ & $ -0.1 \pm 0.08  $ & $  1 \pm 1     $ & $ 4 \pm 4 $ \\ \hline
 	 Oval 5 & $ -28.8 \pm 0.3 $ & $  0.4 \pm 0.1   $ & $ -5 \pm 1     $ & $ -16 \pm 2 $ \\ \hline
	 Oval 6 & $ -29.2 \pm 0.3 $ & $  0.49 \pm 0.03 $ & $ -6.1 \pm 0.4 $ & $ -15 \pm 3 $ \\ \hline
	 Oval 7 & $ -28.8 \pm 0.8 $ & $  0.2 \pm 0.3   $ & $ -2 \pm 4     $ & $ -16 \pm 2 $ \\ \hline
	 Oval 8 & $ -26.8 \pm 0.3 $ & $ -0.5 \pm 0.1   $ & $  6 \pm 1     $ & $ 13 \pm 4 $ \\ \hline
	 Oval 9 & $ -28.0 \pm 0.5 $ & $ 0.5 \pm 1     $ & $  -6 \pm 1     $ & $ -6 \pm 4 $ \\ \hline
	 Oval 10 & $ -28.5 \pm 0.5 $ & $ 0.5 \pm 1     $ & $  -6 \pm 1     $ & $ -13 \pm 3 $ \\ \hline
	 Oval forming 1 & $ -29.9 \pm 0.3 $ & $ -0.15 \pm 0.05 $ & $ 0.8 \pm 0.6   $ & $ -5 \pm 4 $ \\ \hline
	 Oval forming 2 & $ -28.1 \pm 0.2 $ & $  0.63 \pm 0.05 $ & $ -7.9 \pm 0.6 $ & $ -7 \pm 3 $ \\ \hline
	 Oval west 1 & $ -28.7 \pm 0.3 $ & $  0.26 \pm 0.09 $ & $ -3 \pm 1     $ & $ -15 \pm 2 $ \\ \hline
	 Oval west 2 & $ -28.6 \pm 0.2 $ & $  0.3 \pm 0.2   $ & $ -4 \pm 2     $ & $ -13 \pm 3 $ \\ \hline
	 South structure & $ -29.7 \pm 0.3 $ & $  0.07 \pm 0.05 $ & $ -0.9 \pm 0.6 $ & $ -9 \pm 3 $ \\ \hline
  \multirow{2}{*}{Spot} & $ -29.1 \pm 0.5 $ & $ 0.64 \pm 0.07 $ & $ -7.9 \pm 0.9 $ & $ -15 \pm 3 $ \\ \cline{2-5}
                    & $ -29.8 \pm 0.4 $ & $ 0.05 \pm 0.04 $ & $ -0.6 \pm 0.5 $ & $ -7 \pm 4 $ \\ \hline
  \multirow{2}{*}{Spot 1} & $ -29.1 \pm 0.2 $ & $ 0.90 \pm 0.02 $ & $ -11.2 \pm 0.2 $ & $ -15 \pm 3 $ \\ \cline{2-5}
                    & $ -30.1 \pm 0.3 $ & $ -0.08 \pm 0.04 $ & $ 1.0 \pm 0.5 $ & $ -1 \pm 3 $ \\ \hline
	 Spot 2 & $ -29.0 \pm 0.2 $ & $ 0.57 \pm 0.08 $ & $ -7 \pm 1      $ & $ -16 \pm 3 $ \\ \hline
	 Spot 3 & $ -28.6 \pm 0.2 $ & $ 0.8 \pm 0.2   $ & $ -10 \pm 2     $ & $ -13 \pm 3 $ \\ \hline
	 Spot 4 & $ -28.6 \pm 0.4 $ & $ 1.11 \pm 0.07 $ & $ -13.8 \pm 0.9 $ & $ -13 \pm 3 $ \\ \hline
  \multirow{2}{*}{Spot 5} & $ -29.6 \pm 0.5 $ & $ 0.55 \pm 0.04 $ & $ -6.8 \pm 0.5 $ & $ -10 \pm 3 $ \\ \cline{2-5}
                    & $ -29.9 \pm 0.4 $ & $ -0.09 \pm 0.04 $ & $ 1.1 \pm 0.5 $ & $ -5 \pm 4 $ \\ \hline
	 Spot 6 & $ -30.0 \pm 0.3 $ & $ -0.1 \pm 0.4 $ & $ 1 \pm 5        $ & $ -3 \pm 4 $ \\ \hline
	 Spot 7 & $ -29.9 \pm 0.4 $ & $ 2 \pm 2       $ & $ -25 \pm 25    $ & $ -5 \pm 4 $ \\ \hline
	 Spot 8 & $ -30.0 \pm 0.1 $ & $ -2 \pm 2       $ & $ 25 \pm 25    $ & $ -3 \pm 4 $ \\ \hline
	 West fragment & $ -29.7 \pm 0.3 $ & $ 0.4 \pm 0.1   $ & $ -5 \pm 1      $ & $ -9 \pm 3 $ \\ \hline
\end{xtabular}
\end{footnotesize}

\phantomsection
\addcontentsline{toc}{section}{References}

\bibliography{mybibfile}

\end{document}